\documentclass[aps,prl,10pt,twocolumn,showpacs,superscriptaddress,groupedaddress,floatfix]{revtex4-1}

\usepackage[english]{babel}

\usepackage[colorlinks]{hyperref}
\usepackage{mathtools}
\usepackage{latexsym}
\usepackage{graphicx}
\usepackage{float}
\usepackage{bm}
\usepackage{amssymb}
\usepackage{amsmath}
\usepackage{mathtools}
\usepackage[space]{grffile}
\usepackage[usenames,dvipsnames]{xcolor}
\usepackage{comment}
\usepackage{slashed}
\usepackage{physics}
\usepackage[colorinlistoftodos]{todonotes}

\DeclareSymbolFont{matha}{OML}{txmi}{m}{it}
\DeclareMathSymbol{\varv}{\mathord}{matha}{118}

\usepackage[normalem]{ulem}

\def\XXint#1#2#3{{\setbox0=\hbox{$#1{#2#3}{\int}$}
     \vcenter{\hbox{$#2#3$}}\kern-.5\wd0}}

\newcommand{\ie}{\textit{i.e. }}
\newcommand{\eg}{\textit{e.g.  }}
\newcommand{\im}{\mathrm{Im}\,}
\newcommand{\re}{\mathrm{Re}\,}
\newcommand{\der}{\partial}

\newcommand{\ii}{\mathrm{i}}
\newcommand{\niceref}[1] {Eq.~(\ref{#1})}

\newcommand{\be}{\begin{equation}}
\newcommand{\ee}{\end{equation}}
\newcommand{\bea}{\begin{align}}
\newcommand{\eea}{\end{align}}

\newcommand{\phd}{{\phantom{\dagger}}}

\begin{document}

\title{
Phase transitions in full counting statistics of free fermions and directed polymers}
\author{James S. Pallister}
\affiliation{School of Physics and Astronomy, University of Birmingham, Edgbaston, Birmingham, B15 2TT, UK }

\author{Samuel H. Pickering}
\affiliation{School of Physics and Astronomy, University of Birmingham, Edgbaston, Birmingham, B15 2TT, UK }

\author{Dimitri M. Gangardt}
\affiliation{School of Physics and Astronomy, University of Birmingham, Edgbaston, Birmingham, B15 2TT, UK }

\author{Alexander G. Abanov }
\affiliation{Department of Physics and Astronomy, Stony Brook University,
Stony Brook, NY 11794, USA
}

\date{\today}

\begin{abstract}
We consider directed polymers in 1+1 spatial dimension under action of an external repulsive  potential  along a line.  Using  
the exact  mapping onto imaginary time evolution of  free fermions we find that for sufficiently strong potential the system of polymers undergoes a continuous configurational phase transition. The transition corresponds to  merging empty regions in the dominant limit shape.

\end{abstract}

\maketitle


\paragraph{Introduction}
Limit shapes, \ie the emergence of  dominant configurations with sharp boundaries in fluctuating statistical systems, are spectacular phenomena playing an important role in studying universal macroscopic  behaviour in statistical physics \cite{kenyon2009lectures,Stephan_2021}.  
Recently, this phenomenon was discovered in directed polymers or vicious walkers \cite{Rocklin_2012} where it was shown that the strong mutual repulsion between polymers gives rise to emergent long-range entropic forces 
that combine with rigid topological constraints to produce sharp features of the coarse-grained polymer density. 

In quantum many-body systems limit shape configurations were reported in studies of the so-called Emptiness Formation Probability (EFP) \cite{korepin1993quantum, abanov_emptiness_2003,Franchini2005b,abanov-hydro,stephan_emptiness_2014}, the probability $P_R(0)$ that a macroscopic (on the scale of inter-particle separation) interval $I=[-R/2,R/2]$  is completely devoid of particles. The quantum hydrodynamic approach \cite{abanov-hydro} 
has elucidated the relationship between EFP and a limit shape -- the optimal density profile with an astroid-shaped boundary in euclidean space-time. In recent works similar limit shapes were found for weakly interacting bosonic liquids  \cite{yeh_emptiness_2020}  and polytropic gases \cite{yeh_emptiness_2022}. 

It is known since the work of de Gennes \cite{gennes_soluble_1968}  that non-crossing directed polymers  are exactly mapped to   world-lines of nonineracting  fermions propagating in imaginary time.  
Thus, the combination of the Pauli Principle and the emptiness constraint leads naturally to essentially the same limit shape phenomenon.

EFP is a particular case of Full Counting Statistics (FCS), defined via the generating function 
\begin{align}\label{eq:chi}
    \chi_R(s) = \expval{e^{-sN_R}} = \sum_{N=0}^\infty P_R(N) e^{-s N} \, .
\end{align}
FCS is defined as the probability $P_R(N)$ that the number of polymers  crossing the interval $N_R=N$. 
EFP is obtained by taking $s\to +\infty$ limit in \eqref{eq:chi}.

FCS was extensively studied in the context of quantum noise 
for mesoscopic condensed matter 
\cite{levitov_charge-transport_1992,nazarov_quantum_2003}, ultra-cold atomic systems \cite{polkovnikov_interference_2006,gritsev_full_2006,Arzamasovs_2019}, and one-dimensional magnetism \cite{eisler_magnetization_2003,lamacraft_order_2008,Ivanov2013b}. The cumulants of FCS are universally related to  quantum entanglement entropy \cite{klich_quantum_2009,Calabrese_2011,Vicarientangle,calabrese_random_2015}.
Recently, the generating function \niceref{eq:chi} appeared  naturally in the context of continuously measured quantum systems \cite{garratt_measurements_2023}. 
It was proposed in Ref.~\cite{Ivanov2013b} to use analytical properties of the FCS generating function  to characterise the thermodynamic phases and to probe nonlocal correlations in many-body quantum systems.

In these works FCS was studied in the regime of \emph{typical} fluctuations where the corresponding probability distribution $P_R(N)$ is almost Gaussian with higher cumulants supressed by inverse powers of $R$  
\cite{ivanov2013counting, stephan_emptiness_2014}. 
In contrast,  EFP represents a rare event in which the density of particles in the interval deviates macroscopically from its average value $\nu_\infty=\expval{N_R}/R$. The corresponding optimal instanton configuration in 1+1 dimensions has a peculiar astroid shape with typical dimensions $R\times R$. 
This motivates our study of FCS in the large deviation form 
\begin{align}\label{eq:ldscaling}
    \chi_R(s) \sim e^{-R^2 f(s/R)},
\end{align}
in terms  of the rescaled parameter $\alpha=s/R$. In this Letter  we show  that the transition from typical to rare fluctuations is accompanied by a structural change of the dominant limit shape.

\paragraph{Model.}
We consider a statistical ensemble of non-intersecting paths  $\{x_i (\tau)\}$, $x_i (\tau)< x_{i+1}(\tau)$ representing  directed polymers. The polymers are parameterized by the continuous longitudinal  coordinate $\tau, \tau\in[-\beta/2,\beta/2]$ and lateral coordinates $x_i\in \mathbb{Z}$ which take values on a one-dimensional lattice and obey $x_i (\beta/2)=x_i(-\beta/2) $.
Their density is $\nu(x,\tau) = \sum_i \delta_{x,x_i(\tau)}$, where $\delta_{x,y}$ is the Kronecker delta. In terms of the polymers,  FCS  is the distribution  of  the observable $N_R$ equal  to the number of polymers with $|x_i(0)|<R/2$.  

Directed polymers can be mapped onto the worldlines of non-interacting fermions hopping between nearest neighbor sites on a one dimensional lattice propagating in imaginary time $\tau$ \cite{gennes_soluble_1968,fisher_walks_1983,Rocklin_2012}. Their evolution is governed by the tight-binding Hamiltonian $H$ given by Eq.~\eqref{eq:ham} below. In terms of this Hamiltonian, the generating  function of FCS can then be represented as the many-body quantum expectation value
\begin{align}\label{eq:GC}
    \chi_R(s) = \expval{e^{-s N_R}}= \frac{\mathrm{Tr} \left(e^{-s N_R} e^{-\beta H} \right)}{\mathrm{Tr} \left(e^{-\beta H}\right)}\, .
\end{align}
which is the main object of our study.

\begin{figure}[t]
    
\centering
 \includegraphics[width=0.213\textwidth]{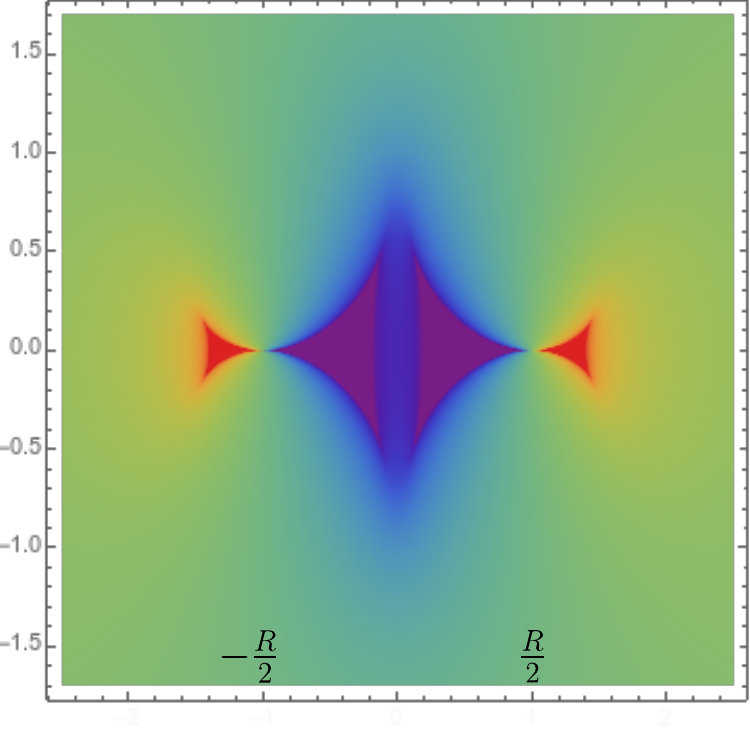}
 \includegraphics[width=0.247\textwidth]{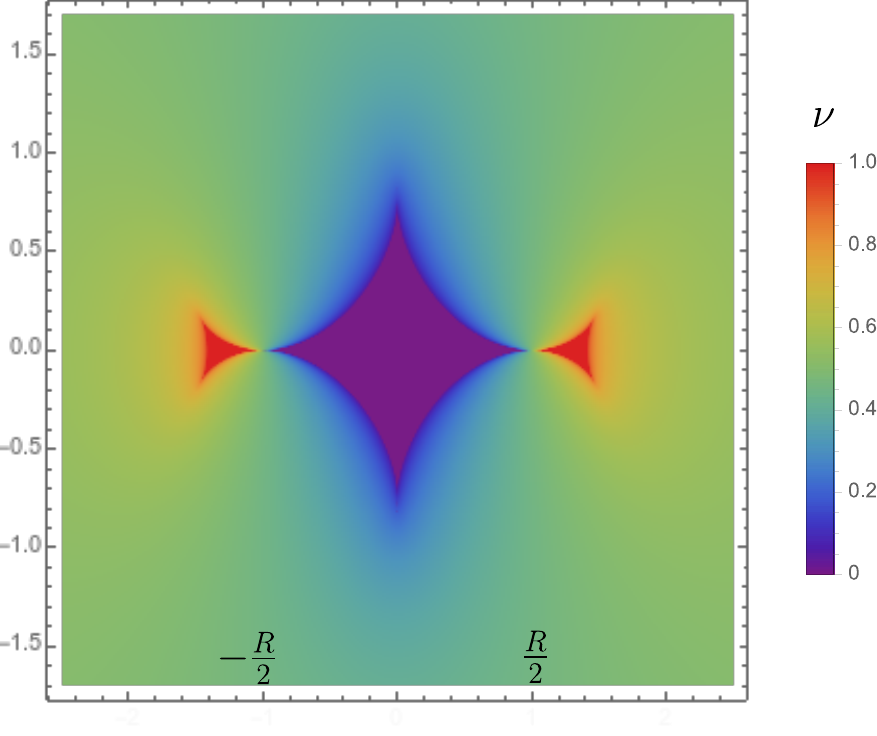}
    \includegraphics[width=0.22
\textwidth]{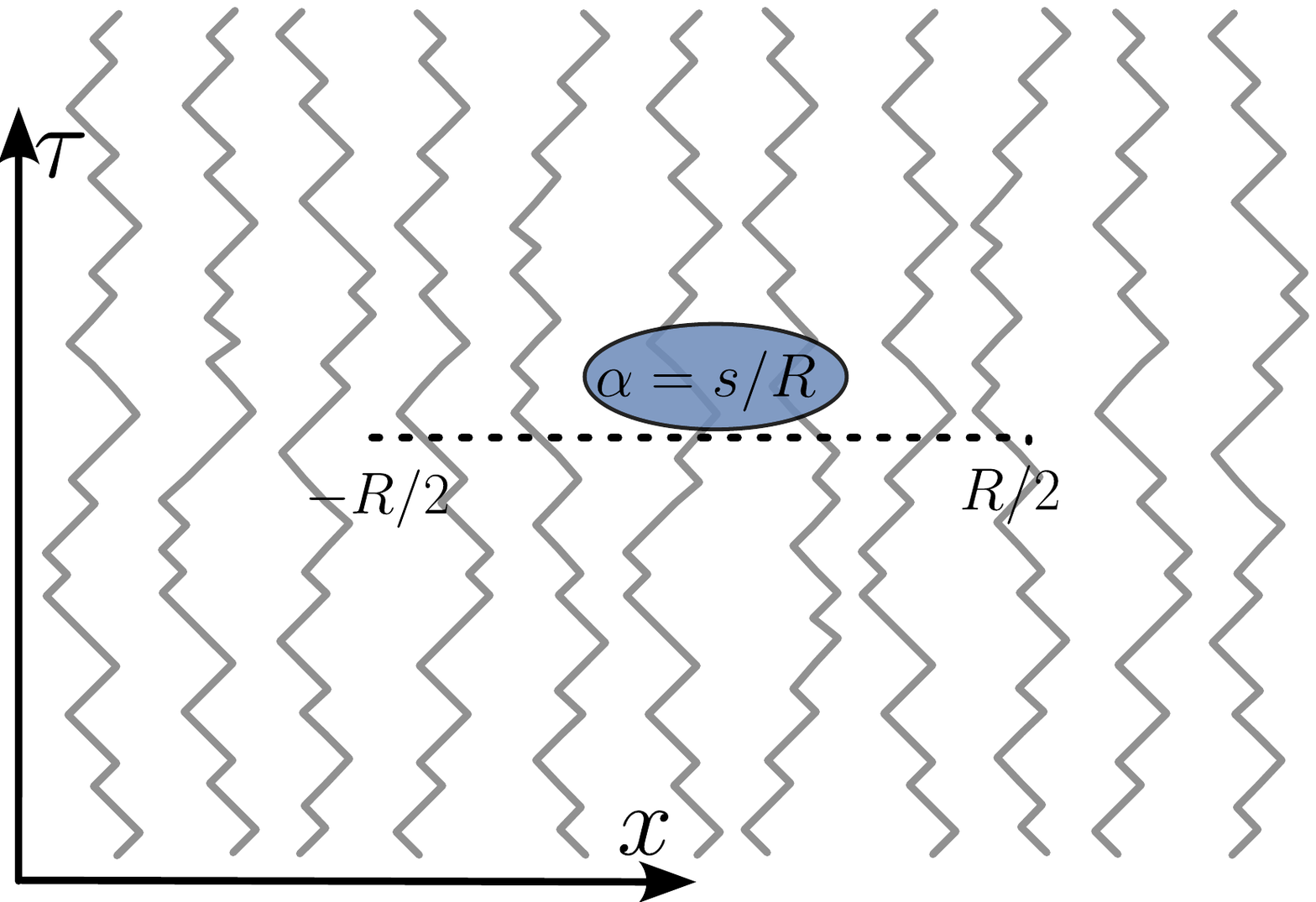}
  \includegraphics[width=0.22\textwidth]{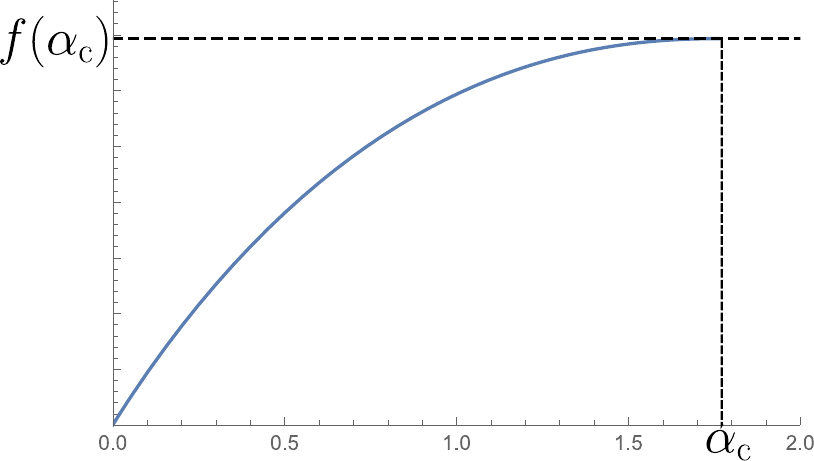}

    \caption{ Top: the dominant density configuration $\nu(x,\tau)$ is shown for $\alpha<\alpha_\mathrm{c}$ (left) and $\alpha\ge \alpha_\mathrm{c}$ (right) is shown for   $\nu_\infty=1/2$. Bottom left: a sketch of polymers' configuration. Each polymer passing through the interval $I=[-R/2,R/2]$ is penalised by weight $e^{-\alpha R}$. Bottom right: the rate function $f(\alpha)$, see Eq.~\eqref{eq:ldscaling}.  }
    \label{fig:trans}
\end{figure}

\paragraph{Results.} The main result  is  
a configurational transition of directed polymers at the critical value $\alpha_c$, illustrated in Fig.~\ref{fig:trans} by plotting the typical density
$\nu(x,\tau)$. The latter is obtained  by solving 
\begin{align}\label{eq:char}
    x+\ii \tau \sin k  = Rg(e^{\ii k}),
\end{align}
for the real part of the complex momentum  $k(x,\tau )=\pi\nu+\ii\upsilon$ at fixed $x,\tau$.  The function $g(z)$ is given by 
\begin{align}\label{eq:gz}
    g(z) = \frac{1}{2}\sqrt{
    \frac{(z-e^{\ii k_0})(z-e^{-\ii k_0})}{(z-e^{\ii k_F})(z-e^{-\ii k_F})}}\, .
\end{align}
Here $k_F=\pi\nu_\infty$ is proportional to  the asymptotic density $\nu_\infty$ at $x,\tau\to \infty$ and  $k_0$ is controlled by $\alpha$  via 
\begin{align}\label{eq:fixalpha}
    \alpha =2 \re \int_{-1}^{1} \frac{g(z)\dd z}{z}\, ,\qquad k_0>0 
\end{align}
and is related by $k_0=\pi\nu(0,0)$ to the density in the center of the interval $I$. 

For moderate $\alpha$, the optimal polymer density consists  of two deltoid-shaped empty ($\nu=0$) and two deltoid-shaped full ($\nu=1$) regions, as shown in the left panel of 
Fig.~\ref{fig:trans}.  The  number of polymers in the interval $I$ is macroscopically depleted with average density
\begin{align}\label{eq:nuIint}
    \nu_I =f'(\alpha)= \frac{1}{\pi}\im \int_{e^{-\ii k_F}}^{e^{\ii k_F}}\frac{g(z)\dd z}{z}\, . 
\end{align}

The value $k_0  =0$  marks the  onset of the transition to a macroscopically empty interval, $\nu_I=0$. At this point 
the integral \eqref{eq:fixalpha}
reaches its maximum value $\alpha_\mathrm{c}$ given by
\begin{align}\label{eq:alphac}
\cosh \frac{\alpha_c}{2} \cos\frac{k
_F}{2} =1 \, .
\end{align}
The right panel of Fig.~\ref{fig:trans} shows
the empty regions having merged 
into one astroid-shaped empty region which includes the whole interval $I$.  This situation, where polymers avoid the interval completely, persists for $\alpha>\alpha_c$ with the constant 
rate function  given by the  EFP result of Ref.~\cite{abanov-hydro},
\begin{align}\label{eq:falphac}
f(\alpha) =f(\alpha_\mathrm{c}) =-\log\cos\frac{k_F}{2}\, ,\qquad \alpha>\alpha_\mathrm{c} .      
 \end{align}

Just below the transition $\alpha\lesssim\alpha_\mathrm{c}$
the behaviour of the rate function is found to be
\begin{align}\label{eq:critf}
    f(\alpha_\mathrm{c}-\delta)=f(\alpha_\mathrm{c})
    +\frac{\delta^2}{2\log \delta}\, ,\qquad \delta\ll 1  .  
\end{align}
Thus the transition is continuous, featuring a divergent third derivative of the rate function similar to the transitions reported earlier in \cite{kazakov1988recent,majumdar_index_2009,Marino_vari}.  

\paragraph{Derivation.}
To generate the configurations of the polymers via fermionic evolution  we use the  tight-binding second quantized Hamiltonian  
\begin{align}\label{eq:ham}
    H=
    \sum_{xy} (h)_{xy} c_x^\dagger c^\phd_y 
\end{align}
 expressed via  creation/annihilation operators $c^\dagger_x,c^\phd_x$ obeying standard anti-commutation relations $\{c^\phd_x,c^\dagger_{x'}\} = \delta_{xx'}$.   The nearest neighbor hopping  matrix
\begin{align}\label{eq:ham1}
    h_{xy}=\int_{-\pi}^\pi\frac{\dd k}{2\pi} 
    e^{\ii k (x-y)} \left(-\cos k -\mu \right),
\end{align}
contains the tight-binding dispersion  $\varepsilon(k)=-\cos k$ and a chemical potential $\mu$ fixing the equilibrium density.  The operator of  particle number  in an interval $I$ is
\begin{align}\label{eq:n}
    N_R = \sum_{xy} (n_I)_{xy} c_x^\dagger c^\phd_y =\sum_{x\in I} c^\dagger_{x} c^\phd_x \, ,
\end{align}
where $(n_I)_{xy} =\delta_{xy}$, for $x\in I$ and zero otherwise. It is  the one-particle projector, $n_I=n_I^2$,  on the interval $I$. Due to the particle-hole symmetry, the  FCS obeys $\chi_R(s,\beta,\mu) = e^{-sR} \chi_R (-s,\beta,-\mu)$ so in what follows we focus on $s\ge 0$.

To find the generating function \niceref{eq:GC}
we  express, following \cite{klich2002counting},  the traces in many-particle Fock space as  determinants   of operators in one-particle space: 
\begin{align}
    \chi_R(s) = \frac{ \mathrm{det}\left(1 + e^{- s n_I} e^{-\beta h}\right)}{\mathrm{det}\left(1 + e^{-\beta h}\right)}.
 \label{chiR14}
\end{align}
The fact that  $n_I$ projects into $R$ lattice sites allows us to rewrite $\chi_R(s)$ as a determinant of  $R\times R$ Toeplitz matrix
with the  elements 
\begin{align}\label{eq:txy}
 \left(1+(e^{-s}-1)n_F\right)_{xy}=\int_{-\pi}^\pi\frac{\dd k}{2\pi} 
e^{\ii k (x-y)} e^{-sv(k)}\, ,
\end{align}
where $n_F = (1+e^{\beta h})^{-1}$ is the Fermi occupation operator.

We consider the zero temperature limit, $\beta\to \infty$ corresponding to infinitely long polymers. In this limit $n_F$ becomes the projector on the single interval 
in the momentum space, $|k|<k_F$ with
$\mu =-\cos k_F$. 
The so-called  Toeplitz symbol $e^{-s v(k)}$ becomes singular since  
\begin{align} \label{eq:vk}
v(k) = -\frac{1}{s} \log\frac{1+e^{-s} e^{-\beta h(k)}}{1+e^{-\beta h(k)}} \to   \theta(k_F- \abs{k})\, .  
\end{align}
Although the theory of Toeplitz determinants with singular symbols is well developed, see, \textit{e.g.} Ref.~\cite{deift_asymptotics_2011},  it is not directly applicable in our case where the parameter $s$ scales with the size of the matrix $R$. Here we use an alternative approach based on the exact representation of the generating function  as the  $R$-dimensional integral,
\begin{align}\label{eq:chifermi}
    \chi_R(s) &= \frac{1}{R!}
    \int  \frac{\dd^R k}{(2\pi)^R}  \left|\Delta(e^{\ii k
})\right|^2 \prod_{l=1}^R e^{-s v(k_l)},
\end{align}
which is a configurational integral of a classical two-dimensional Coulomb gas of $R$ charges \cite{SM}.
In addition to a repulsive interaction given by the logarithm of Vandermonde determinant, $\Delta(e^{\ii k})=\prod_{1\le m<n\le R}(e^{\ii k_n}-e^{\ii k_m})$, the charges on the  interval $|k| < k_F$ are subject to a constant electrostatic repulsive potential of magnitude 
$s=\alpha R$.  This scaling makes   
the potential comparable to the interparticle repulsion, both terms being proportional to $R^2$ justifying  the scaling in  Eq.~\eqref{eq:ldscaling}. 

Due to the presence of large parameter $R$ the integral \eqref{eq:chifermi} is dominated by a stationary configuration of charges. Their positions, $z_l=e^{\ii k_l}$, are encoded in the poles of the resolvent   
\begin{align}
    g(z)+\frac{1}{2} = \frac{1}{R}\sum_{l=1}^R \frac{z}{z-e^{\ii k_l}} = \int_{-\pi}^\pi\frac{\dd k}{2\pi} \frac{z\sigma (k)}{z-e^{\ii k}}\, .
\end{align}
Here $\sigma(k) = \frac{1}{R}\sum_{l=1}^R \delta(k-k_l)$ is the density of the poles, 
which  becomes the smooth function, \niceref{eq:gz}, for $R\to\infty$. In this limit the charges are distributed along the branch cuts of $g(z)$, those being the two disjoint arcs of unit circle $z=e^{\ii k}$, $|k|<k_0$  and $k_F<|k|<\pi$. 
\begin{figure}[t]
\centering
  \includegraphics[width=0.22\textwidth,trim=0 -5.3cm 0 0]
  {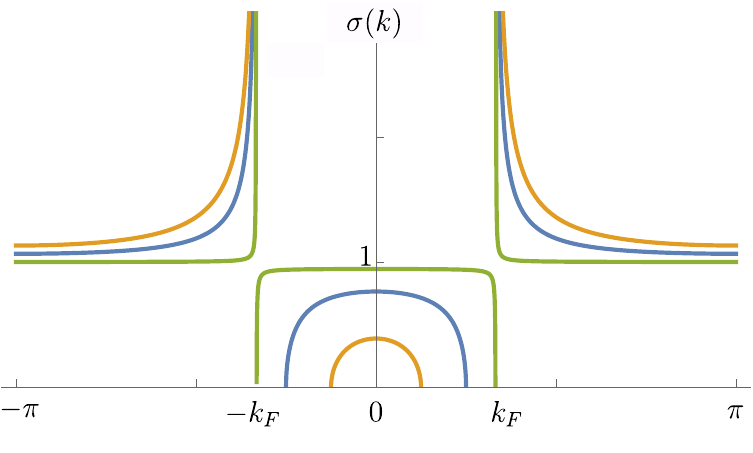}
  \includegraphics[width=0.22\textwidth]
  {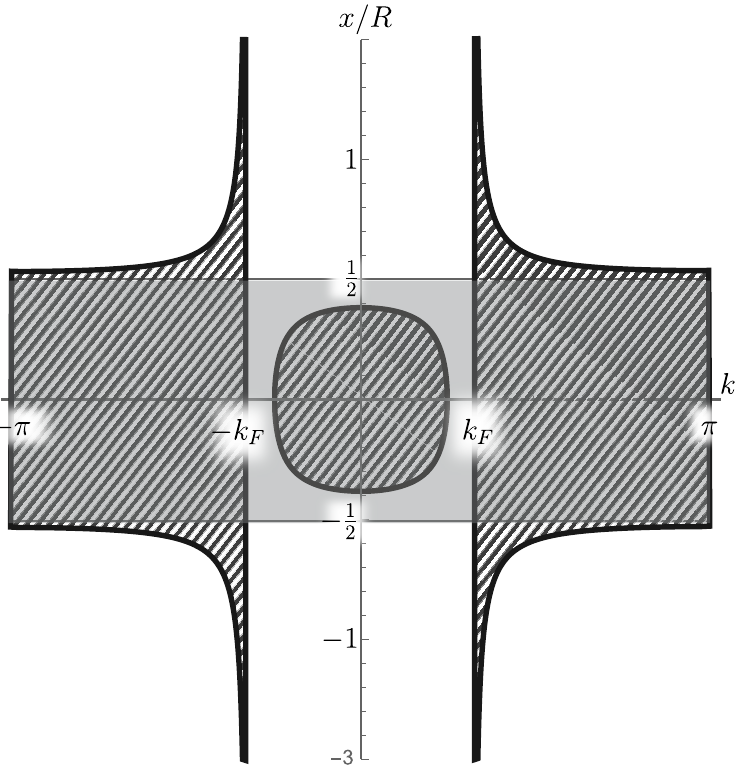}
    \caption{Left: Quasiparticle momentum distribution corresponding to the solution of Coulomb Gas problem Eq.~\eqref{eq:sigmak} for $k_F/\pi=1/3$ and  $\alpha$ corresponding to $k_0/\pi = 0.332,0.25,0.125$.  Right: quasiparticle Wigner function $f (x,k)$ for  $k_0/\pi = 0.25$. The striped areas correspond to regions where $f(x,k)=1$.  
    The light grey rectangle represents regions where the quasiparticle vacuum ($s=0$) Wigner function $f^{\mathrm{vac}}_\mathrm{qp} (x,k)=1$.}
    \label{fig:distributions}
\end{figure}

Introducing the complex potential, $g(z)+1/2 = z\der_z\varphi(z) $,  the stationarity condition can be expressed as the electrostatic equilibrium of the charges: the real part of $\varphi(z)$ is constant along each of  the branch cuts. The parameter $\alpha$ then plays the role of the voltage between the cuts, hence the condition \eqref{eq:fixalpha} fixing  the parameter  $\nu_0$. 
Conversely, the imaginary part of the potential counts the number of charges and leads to 
Eq.~\eqref{eq:nuIint}. The density corresponding to the resolvent $g(z)$ in Eq.~\eqref{eq:gz} is 
\begin{align}\label{eq:sigmak}
    \sigma(k) = 2\re g(e^{\ii k})=\sqrt{\frac{\sin^2\frac{k}{2}-\sin^2\frac{k_0}{2}}{\sin^2\frac{k}{2}-\sin^2\frac{k_F}{2}} }\, .
\end{align}
and is shown in the left panel of Fig.~\ref{fig:distributions}\,. The calculations leading to these results are standard and can be found in Supplemental Material \cite{SM}.

\paragraph{Hydrodynamic interpretation}
The  solution of the  Coulomb Gas model 
contains information about  the \emph{state} of the system at $\tau=0$ which, using the hydrodynamic evolution 
of fermions, allows one to extract  full information about the real space configurations of constrained directed polymers. 
This can be traced back to interpreting Eq.~\eqref{eq:GC} as an imaginary time path integral with the source term proportional to the counting parameter $s$ acting at $\tau=0$ and affecting $R$ sites of the interval $I$. 

The configurations contributing to the generating function  \niceref{eq:chi} can be parametrised by degrees of freedom of $R$ fermionic quasiparticles propagating in imaginary time with $R\sigma(k)$ playing the role of their momentum distribution at $\tau=0$. 
 
It is well known 
\cite{abanov-hydro} that the 
large scale dynamics of such  
fermionic quasiparticles  is  captured by  
the hydrodynamic approach. In this approach one
introduces two hydrodynamic fields, the density $\nu(x,\tau)$ and velocity $\upsilon(x,\tau)$  combined into complex momentum $k(x,\tau)=\pi\nu+\ii \upsilon$.  The complex momenta 
satisfy the complex Burger's equation
\begin{align}\label{eq:Burger}
    \ii\der_\tau k + \der_x\varepsilon(k) = 0 \, , \qquad\tau\neq 0\, 
\end{align}
equivalent to hydrodynamic continuity and Euler equations after separation into real and imaginary parts. 

As a result of time-reversal symmetry $k(x,-\tau) = \bar{k}(x,\tau)$ the density $\nu(x,\tau)$ and velocity $\upsilon(x,\tau)$ fields are even/odd functions of $\tau$ correspondingly. The initial condition 
$k_0(x)= k(x,0)$ is, therefore, real.
The inverse function $x_0(k)$ such that $x_0 (k_0(x))=x$
parametrises  the implicit solution to \niceref{eq:Burger}
\begin{align}\label{eq:Burgers_sol}
    x+\ii \varepsilon'(k) \tau = x_0(k)\, .
\end{align}
Below we argue that the right hand side is given by the rescaled solution of the Coulomb gas problem as 
\begin{align}\label{eq:xsigma}
x_0(k)  = R\sigma(k)/2 = Rg(e^{\ii k})\, .    
\end{align}

Indeed, for $s=0$ the solutions of hydrodynamic Eq.~\eqref{eq:Burger} are given by the constant configuration  $k(x,\tau) = \bar{k} (x,\tau) = k_F= \pi \nu_\infty $.
For $s >0$ the density is non-uniform and
we write  
 \begin{align}
     \nu (x,0) -\nu_\infty = \int\frac{\dd k}{2\pi} \left(f (x,k) - f^{\mathrm{vac}}(x,k)\right)\, .
 \end{align}
Here we represented the density fluctuation as a sum of contributions from different momenta, using the semiclassical Wigner function $f(x,k)$.
The  vacuum contribution  $f^{\mathrm{vac}}(x,k)$ corresponding to uniform density $\nu_\infty$ at $s=0$, is subtracted as $\nu-\nu_\infty=0$ for  $s=0$. 
The function $f(x,k)$ has the form of a local Fermi distribution, $f(x,k) = 1$ for  $|k|<k_0 (x)$  and zero outside this interval, see Fig.~\ref{fig:distributions}. The branches  $\pm k_0 (x)$ are thus the  local ``Fermi momenta'' of the fermionic particles. 

The same boundary can be represented 
using  the inverse  function $x_0(k)$. 
In terms of the Wigner function we have
\begin{align}\label{eq:x0}
 2 x_{0}(k)=\sum_x[ f (x,k)-f^\mathrm{vac} (x,k) + f_\mathrm{qp}^\mathrm{vac} (x,k)]  \, .
\end{align}
Here we added the Wigner function $f_\mathrm{qp}^\mathrm{vac} (x,k)=\theta(R/2-|x|)$ representing the vacuum ($s=0$) distribution of \emph{quasiparticles} as shown in Fig.~\ref{fig:distributions}.
The sum in the r.h.s. of Eq.~\eqref{eq:x0} is nothing but the momentum distribution of the quasiparticles $R\sigma(k)$.  
This  establishes the important link,    
Eq.~\eqref{eq:xsigma}, between the Coulomb Gas charge density and the fluid dynamics.

The solution $k_0(x)$ of  Eq.~\eqref{eq:x0} at $\tau=0$, shows  that  
for $0<\alpha<\alpha_\mathrm{c}$ there is a non-zero density in some interval around $x=0$.
As one moves towards the ends of the interval the density drops to zero at $\pm x_- = R g(1)$. At $x=\pm R/2$ the density jumps to unity and stays saturated up to $\pm x_+ =R g(-1) $. Beyond this point the density decays to its asymptotic value $\nu_\infty = k_F/\pi$.

Solving \niceref{eq:Burgers_sol} away from $\tau=0$ provides the full space-time profile of density and velocity in terms of the complex function $k(x,\tau)$. 
For $0<\alpha<\alpha_\mathrm{c}$ there is always a central fluctuating region between the deltoid-shaped frozen regions which evolve from the frozen intervals with $\nu=0,1$ at $\tau=0$. The boundary of the frozen regions is given in the parametric form
\begin{align}
    x/R&= G_\pm(v) -  \tanh{v} \,G'_\pm (v),  \\
    \tau/R &= \mp G'_\pm (v)/\cosh v \, ,  
\end{align}
which is obtained by substituting $k_+=\ii v, k_-=\pi+\ii v $ into Eq.~\eqref{eq:char} and defining $G_\pm (v) = g(e^{\ii k_\pm}) =g(\pm e^{-v}) $ for the boundary with empty/full frozen region.

When $\alpha$ is increased from zero the   parameter $\nu_0$ decreases monotonically from $\nu_0=\nu_\infty$.
The transition corresponds to disappearance of the cut, $\nu_0=0$, at the critical value $\alpha_\mathrm{c}$. Calculation of $\alpha_\mathrm{c}$, Eq.~\eqref{eq:alphac} and the corresponding value of free energy, Eq.~\eqref{eq:falphac} are outlined in Supplemental Material \cite{SM}.

To establish the critical behaviour close to the transition,  $\nu_0\to 0$,   we  consider Eqs.~\eqref{eq:fixalpha} and \eqref{eq:nuIint}. The density $\sigma(k)$ is zero everywhere in $[-k_F,k_F]$ except for a small interval $[-k_0,k_0]$ where it can be approximated by the Wigner semicircle. The standard calculation gives
\begin{align}
    \nu_I =\frac{k_0}{2\sin\frac{k_F}{2}} \int_{-k_0}^{k_0} \frac{\dd k}{2\pi}\sqrt{1-\left(\frac{k}{k_0}\right)^2}  = \frac{k_0^2}{8\sin\frac{k_F}{2}}.
\end{align}

Calculating $\alpha$ is more involved. One can show \cite{SM} that the corresponding integral contains a logarithmic non-analyticity:
\begin{align}
    \alpha-\alpha_\mathrm{c}   =- \frac{k_0^2}{8\sin\frac{k_F}{2}}\left[1+\log\left(\frac{64}{k_0^2}\tan^2\frac{k_F}{2}\right)\right]\, .
\end{align}
Thus in the leading order $\delta=\alpha_\mathrm{c}-\alpha \simeq \nu_I\log(1/\nu_I)$, which can be inverted as $\nu_I = \delta/\log(1/\delta)$. Integrating $f'(\alpha) =\nu_I $ from $\alpha_\mathrm{c}$ to $\alpha =\alpha_\mathrm{c} - \delta$ one gets  Eq.~\eqref{eq:critf}.

\paragraph{Conclusions}
In this work we have computed the full counting statistics of lattice fermions within a macroscopic interval in the previously unexplored regime of large deviations and found the presence of a continuous phase transition between zero and near zero occupancy of the interval. 
The mapping to Coulomb gas makes our findings in agreement with findings of Refs.~\cite{cunden2019third}.

We obtained an analytic form for the rate function associated with the large deviation using a mapping between the FCS generating function and a Coulomb gas model. In contrast to the previous studies \cite{Marino_2014,marino_number_2016,smith_full_2021} the Coulomb Gas model is naturally formulated in momentum space of $R$ quasiparticles leading to the existence of the following duality.
It is convenient to rewrite (\ref{chiR14}) as 
\begin{align}
    \chi_R(s) &= \mathrm{det}\left(1-n_F + e^{- s n_I} n_F\right) 
 \nonumber \\ 
    &= \mathrm{det}\left(1 - (1-e^{-s})n_I n_F\right)\, .
 \label{chiR20}
\end{align}
In the limit of zero temperature, $\beta\to \infty$, the Fermi occupation operator $n_F$ also becomes a projector on the negative energy states and the expression (\ref{chiR20}) is completely symmetric with respect to the interchange $n_I\leftrightarrow n_F$. One has a choice of computing (\ref{chiR20}) in the momentum basis where it becomes the Fredholm determinant with the sine-kernel or in the coordinate basis, where it becomes the Toeplitz determinant of the size $R\times R$. This change corresponds to switching between the original fermions and the quasiparticles used for the hydrodynamic description (see Fig.~\ref{fig:distributions}).

At zero temperature FCS is, as expected, insensitive to the exact form of the fermion dispersion as long as a single interval of  momenta is occupied in the ground state. This has to be contrasted with the space-time configurations of the hydrodynamic fields depending explicitly on the form of the dispersion, see \eg Eq.~\eqref{eq:Burgers_sol}.  Moreover, for a dispersion leading to  crossing of the worldlines of the fermions the hydrodynamic interpretation may even fail as some configurations would have negative Boltzmann weights \cite{Bocini_2021}. These subtleties disappear  in the continuous limit $k_F,k_0\to 0$ with $k_0/k_F$ fixed, in which $\epsilon(k)\sim k^2$ and Galilean symmetry is restored.

Finally, the momentum space representation used here
can be extended to finite temperatures, by modifying the effective potential $v(k)$, Eq.~\eqref{eq:vk} in the Coulomb Gas model Eq.~\eqref{eq:chifermi}.

\paragraph{Acknowledgements}
We thank Dmitri Ivanov for the critical reading of the preliminary draft of this paper and comments. We gratefully acknowledge the hospitality of the Simons Center for Geometry and Physics during the program ``New Directions in far from Equilibrium Integrability and beyond'' during which this work has been completed.
The work of A.G.A. has been supported by the National Science Foundation under Grant NSF DMR-2116767. A.G.A. also acknowledges the support of Rosi and Max Varon fellowship from the Weizmann Institute of Science and the hospitality of the Galileo Galilei Institute for Theoretical Physics. D.M.G.  would like to thank the Institut Henri Poincaré (UAR 839 CNRS-Sorbonne Université) and the LabEx CARMIN (ANR-10-LABX-59-01) for their support.

\section*{References}


\bibliographystyle{apsrev4-1}

\bibliography{FCS_draft/FCS.bib}

\begin{thebibliography}{37}%
\makeatletter
\providecommand \@ifxundefined [1]{%
 \@ifx{#1\undefined}
}%
\providecommand \@ifnum [1]{%
 \ifnum #1\expandafter \@firstoftwo
 \else \expandafter \@secondoftwo
 \fi
}%
\providecommand \@ifx [1]{%
 \ifx #1\expandafter \@firstoftwo
 \else \expandafter \@secondoftwo
 \fi
}%
\providecommand \natexlab [1]{#1}%
\providecommand \enquote  [1]{``#1''}%
\providecommand \bibnamefont  [1]{#1}%
\providecommand \bibfnamefont [1]{#1}%
\providecommand \citenamefont [1]{#1}%
\providecommand \href@noop [0]{\@secondoftwo}%
\providecommand \href [0]{\begingroup \@sanitize@url \@href}%
\providecommand \@href[1]{\@@startlink{#1}\@@href}%
\providecommand \@@href[1]{\endgroup#1\@@endlink}%
\providecommand \@sanitize@url [0]{\catcode `\\12\catcode `\$12\catcode
  `\&12\catcode `\#12\catcode `\^12\catcode `\_12\catcode `\%12\relax}%
\providecommand \@@startlink[1]{}%
\providecommand \@@endlink[0]{}%
\providecommand \url  [0]{\begingroup\@sanitize@url \@url }%
\providecommand \@url [1]{\endgroup\@href {#1}{\urlprefix }}%
\providecommand \urlprefix  [0]{URL }%
\providecommand \Eprint [0]{\href }%
\providecommand \doibase [0]{http://dx.doi.org/}%
\providecommand \selectlanguage [0]{\@gobble}%
\providecommand \bibinfo  [0]{\@secondoftwo}%
\providecommand \bibfield  [0]{\@secondoftwo}%
\providecommand \translation [1]{[#1]}%
\providecommand \BibitemOpen [0]{}%
\providecommand \bibitemStop [0]{}%
\providecommand \bibitemNoStop [0]{.\EOS\space}%
\providecommand \EOS [0]{\spacefactor3000\relax}%
\providecommand \BibitemShut  [1]{\csname bibitem#1\endcsname}%
\let\auto@bib@innerbib\@empty
\bibitem [{\citenamefont {Kenyon}(2009)}]{kenyon2009lectures}%
  \BibitemOpen
  \bibfield  {author} {\bibinfo {author} {\bibfnamefont {R.}~\bibnamefont
  {Kenyon}},\ }\href@noop {} {\enquote {\bibinfo {title} {Lectures on
  dimers},}\ } (\bibinfo {year} {2009}),\ \Eprint
  {http://arxiv.org/abs/0910.3129} {arXiv:0910.3129} \BibitemShut {NoStop}%
\bibitem [{\citenamefont {Stéphan}(2021)}]{Stephan_2021}%
  \BibitemOpen
  \bibfield  {author} {\bibinfo {author} {\bibfnamefont {J.-M.}\ \bibnamefont
  {Stéphan}},\ }\href {\doibase 10.21468/SciPostPhysLectNotes.26} {\bibfield
  {journal} {\bibinfo  {journal} {SciPost Phys. Lect. Notes}\ ,\ \bibinfo
  {pages} {26}} (\bibinfo {year} {2021})}\BibitemShut {NoStop}%
\bibitem [{\citenamefont {Rocklin}\ \emph {et~al.}(2012)\citenamefont
  {Rocklin}, \citenamefont {Tan},\ and\ \citenamefont
  {Goldbart}}]{Rocklin_2012}%
  \BibitemOpen
  \bibfield  {author} {\bibinfo {author} {\bibfnamefont {D.~Z.}\ \bibnamefont
  {Rocklin}}, \bibinfo {author} {\bibfnamefont {S.}~\bibnamefont {Tan}}, \ and\
  \bibinfo {author} {\bibfnamefont {P.~M.}\ \bibnamefont {Goldbart}},\ }\href
  {\doibase 10.1103/PhysRevB.86.165421} {\bibfield  {journal} {\bibinfo
  {journal} {Phys. Rev. B}\ }\textbf {\bibinfo {volume} {86}},\ \bibinfo
  {pages} {165421} (\bibinfo {year} {2012})}\BibitemShut {NoStop}%
\bibitem [{\citenamefont {Korepin}\ \emph {et~al.}(1993)\citenamefont
  {Korepin}, \citenamefont {Izergin},\ and\ \citenamefont
  {Bogoliubov}}]{korepin1993quantum}%
  \BibitemOpen
  \bibfield  {author} {\bibinfo {author} {\bibfnamefont {V.~E.}\ \bibnamefont
  {Korepin}}, \bibinfo {author} {\bibfnamefont {A.~G.}\ \bibnamefont
  {Izergin}}, \ and\ \bibinfo {author} {\bibfnamefont {N.~M.}\ \bibnamefont
  {Bogoliubov}},\ }\href@noop {} {\enquote {\bibinfo {title} {Quantum inverse
  scattering method and correlation functions},}\ } (\bibinfo {year} {1993}),\
  \Eprint {http://arxiv.org/abs/cond-mat/9301031} {arXiv:cond-mat/9301031}
  \BibitemShut {NoStop}%
\bibitem [{\citenamefont {Abanov}\ and\ \citenamefont
  {Franchini}(2003)}]{abanov_emptiness_2003}%
  \BibitemOpen
  \bibfield  {author} {\bibinfo {author} {\bibfnamefont {A.~G.}\ \bibnamefont
  {Abanov}}\ and\ \bibinfo {author} {\bibfnamefont {F.}~\bibnamefont
  {Franchini}},\ }\href {\doibase 10.1016/j.physleta.2003.07.009} {\bibfield
  {journal} {\bibinfo  {journal} {Phys. Lett. A}\ }\textbf {\bibinfo {volume}
  {316}},\ \bibinfo {pages} {342} (\bibinfo {year} {2003})}\BibitemShut
  {NoStop}%
\bibitem [{\citenamefont {Franchini}\ and\ \citenamefont
  {Abanov}(2005)}]{Franchini2005b}%
  \BibitemOpen
  \bibfield  {author} {\bibinfo {author} {\bibfnamefont {F.}~\bibnamefont
  {Franchini}}\ and\ \bibinfo {author} {\bibfnamefont {A.~G.}\ \bibnamefont
  {Abanov}},\ }\href {\doibase 10.1088/0305-4470/38/23/002} {\bibfield
  {journal} {\bibinfo  {journal} {J. Phys. A}\ }\textbf {\bibinfo {volume}
  {38}},\ \bibinfo {pages} {5069} (\bibinfo {year} {2005})}\BibitemShut
  {NoStop}%
\bibitem [{\citenamefont {Abanov}(2006)}]{abanov-hydro}%
  \BibitemOpen
  \bibfield  {author} {\bibinfo {author} {\bibfnamefont {A.~G.}\ \bibnamefont
  {Abanov}},\ }in\ \href@noop {} {\emph {\bibinfo {booktitle} {Applications of
  Random Matrices in Physics}}},\ \bibinfo {editor} {edited by\ \bibinfo
  {editor} {\bibfnamefont {{\'E}.}~\bibnamefont {Br{\'e}zin}}, \bibinfo
  {editor} {\bibfnamefont {V.}~\bibnamefont {Kazakov}}, \bibinfo {editor}
  {\bibfnamefont {D.}~\bibnamefont {Serban}}, \bibinfo {editor} {\bibfnamefont
  {P.}~\bibnamefont {Wiegmann}}, \ and\ \bibinfo {editor} {\bibfnamefont
  {A.}~\bibnamefont {Zabrodin}}}\ (\bibinfo  {publisher} {Springer
  Netherlands},\ \bibinfo {address} {Dordrecht},\ \bibinfo {year} {2006})\ pp.\
  \bibinfo {pages} {139--161},\ \Eprint {http://arxiv.org/abs/cond-mat/0504307}
  {arXiv:cond-mat/0504307} \BibitemShut {NoStop}%
\bibitem [{\citenamefont {Stéphan}(2014)}]{stephan_emptiness_2014}%
  \BibitemOpen
  \bibfield  {author} {\bibinfo {author} {\bibfnamefont {J.-M.}\ \bibnamefont
  {Stéphan}},\ }\href {\doibase 10.1088/1742-5468/2014/05/P05010} {\bibfield
  {journal} {\bibinfo  {journal} {J. Stat. Mech.}\ }\textbf {\bibinfo {volume}
  {2014}},\ \bibinfo {pages} {P05010} (\bibinfo {year} {2014})}\BibitemShut
  {NoStop}%
\bibitem [{\citenamefont {Yeh}\ and\ \citenamefont
  {Kamenev}(2020)}]{yeh_emptiness_2020}%
  \BibitemOpen
  \bibfield  {author} {\bibinfo {author} {\bibfnamefont {H.-C.}\ \bibnamefont
  {Yeh}}\ and\ \bibinfo {author} {\bibfnamefont {A.}~\bibnamefont {Kamenev}},\
  }\href {\doibase 10.1103/PhysRevA.101.023623} {\bibfield  {journal} {\bibinfo
   {journal} {Phys. Rev. A}\ }\textbf {\bibinfo {volume} {101}},\ \bibinfo
  {pages} {023623} (\bibinfo {year} {2020})}\BibitemShut {NoStop}%
\bibitem [{\citenamefont {Yeh}\ \emph {et~al.}(2022)\citenamefont {Yeh},
  \citenamefont {Gangardt},\ and\ \citenamefont
  {Kamenev}}]{yeh_emptiness_2022}%
  \BibitemOpen
  \bibfield  {author} {\bibinfo {author} {\bibfnamefont {H.-C.}\ \bibnamefont
  {Yeh}}, \bibinfo {author} {\bibfnamefont {D.~M.}\ \bibnamefont {Gangardt}}, \
  and\ \bibinfo {author} {\bibfnamefont {A.}~\bibnamefont {Kamenev}},\ }\href
  {\doibase 10.1088/1751-8121/ac47b1} {\bibfield  {journal} {\bibinfo
  {journal} {J. Phys. A}\ }\textbf {\bibinfo {volume} {55}},\ \bibinfo {pages}
  {064002} (\bibinfo {year} {2022})}\BibitemShut {NoStop}%
\bibitem [{\citenamefont {de~Gennes}(1968)}]{gennes_soluble_1968}%
  \BibitemOpen
  \bibfield  {author} {\bibinfo {author} {\bibfnamefont {P.-G.}\ \bibnamefont
  {de~Gennes}},\ }\href {\doibase 10.1063/1.1669420} {\bibfield  {journal}
  {\bibinfo  {journal} {J. Chem. Phys.}\ }\textbf {\bibinfo {volume} {48}},\
  \bibinfo {pages} {2257} (\bibinfo {year} {1968})}\BibitemShut {NoStop}%
\bibitem [{\citenamefont {Levitov}\ and\ \citenamefont
  {Lesovik}(1992)}]{levitov_charge-transport_1992}%
  \BibitemOpen
  \bibfield  {author} {\bibinfo {author} {\bibfnamefont {L.~S.}\ \bibnamefont
  {Levitov}}\ and\ \bibinfo {author} {\bibfnamefont {G.~B.}\ \bibnamefont
  {Lesovik}},\ }\href@noop {} {\bibfield  {journal} {\bibinfo  {journal} {JETP
  Letters}\ }\textbf {\bibinfo {volume} {55}},\ \bibinfo {pages} {555}
  (\bibinfo {year} {1992})}\BibitemShut {NoStop}%
\bibitem [{\citenamefont {Nazarov}(2003)}]{nazarov_quantum_2003}%
  \BibitemOpen
  \bibfield  {author} {\bibinfo {author} {\bibfnamefont {Y.}~\bibnamefont
  {Nazarov}},\ }\href@noop {} {\emph {\bibinfo {title} {Quantum {Noise} in
  {Mesoscopic} {Physics}}}}\ (\bibinfo  {publisher} {Springer},\ \bibinfo
  {address} {New York},\ \bibinfo {year} {2003})\BibitemShut {NoStop}%
\bibitem [{\citenamefont {Polkovnikov}\ \emph {et~al.}(2006)\citenamefont
  {Polkovnikov}, \citenamefont {Altman},\ and\ \citenamefont
  {Demler}}]{polkovnikov_interference_2006}%
  \BibitemOpen
  \bibfield  {author} {\bibinfo {author} {\bibfnamefont {A.}~\bibnamefont
  {Polkovnikov}}, \bibinfo {author} {\bibfnamefont {E.}~\bibnamefont {Altman}},
  \ and\ \bibinfo {author} {\bibfnamefont {E.}~\bibnamefont {Demler}},\ }\href
  {\doibase 10.1073/pnas.0510276103} {\bibfield  {journal} {\bibinfo  {journal}
  {Proceedings of the National Academy of Sciences}\ }\textbf {\bibinfo
  {volume} {103}},\ \bibinfo {pages} {6125} (\bibinfo {year}
  {2006})}\BibitemShut {NoStop}%
\bibitem [{\citenamefont {Gritsev}\ \emph {et~al.}(2006)\citenamefont
  {Gritsev}, \citenamefont {Altman}, \citenamefont {Demler},\ and\
  \citenamefont {Polkovnikov}}]{gritsev_full_2006}%
  \BibitemOpen
  \bibfield  {author} {\bibinfo {author} {\bibfnamefont {V.}~\bibnamefont
  {Gritsev}}, \bibinfo {author} {\bibfnamefont {E.}~\bibnamefont {Altman}},
  \bibinfo {author} {\bibfnamefont {E.}~\bibnamefont {Demler}}, \ and\ \bibinfo
  {author} {\bibfnamefont {A.}~\bibnamefont {Polkovnikov}},\ }\href {\doibase
  10.1038/nphys410} {\bibfield  {journal} {\bibinfo  {journal} {Nature
  Physics}\ }\textbf {\bibinfo {volume} {2}},\ \bibinfo {pages} {705} (\bibinfo
  {year} {2006})}\BibitemShut {NoStop}%
\bibitem [{\citenamefont {Arzamasovs}\ and\ \citenamefont
  {Gangardt}(2019)}]{Arzamasovs_2019}%
  \BibitemOpen
  \bibfield  {author} {\bibinfo {author} {\bibfnamefont {M.}~\bibnamefont
  {Arzamasovs}}\ and\ \bibinfo {author} {\bibfnamefont {D.~M.}\ \bibnamefont
  {Gangardt}},\ }\href {\doibase 10.1103/PhysRevLett.122.120401} {\bibfield
  {journal} {\bibinfo  {journal} {Physical Review Letters}\ }\textbf {\bibinfo
  {volume} {122}},\ \bibinfo {pages} {120401} (\bibinfo {year}
  {2019})}\BibitemShut {NoStop}%
\bibitem [{\citenamefont {Eisler}\ \emph {et~al.}(2003)\citenamefont {Eisler},
  \citenamefont {Rácz},\ and\ \citenamefont {van
  Wijland}}]{eisler_magnetization_2003}%
  \BibitemOpen
  \bibfield  {author} {\bibinfo {author} {\bibfnamefont {V.}~\bibnamefont
  {Eisler}}, \bibinfo {author} {\bibfnamefont {Z.}~\bibnamefont {Rácz}}, \
  and\ \bibinfo {author} {\bibfnamefont {F.}~\bibnamefont {van Wijland}},\
  }\href {\doibase 10.1103/PhysRevE.67.056129} {\bibfield  {journal} {\bibinfo
  {journal} {Physical Review E}\ }\textbf {\bibinfo {volume} {67}},\ \bibinfo
  {pages} {056129} (\bibinfo {year} {2003})}\BibitemShut {NoStop}%
\bibitem [{\citenamefont {Lamacraft}\ and\ \citenamefont
  {Fendley}(2008)}]{lamacraft_order_2008}%
  \BibitemOpen
  \bibfield  {author} {\bibinfo {author} {\bibfnamefont {A.}~\bibnamefont
  {Lamacraft}}\ and\ \bibinfo {author} {\bibfnamefont {P.}~\bibnamefont
  {Fendley}},\ }\href {\doibase 10.1103/PhysRevLett.100.165706} {\bibfield
  {journal} {\bibinfo  {journal} {Physical Review Letters}\ }\textbf {\bibinfo
  {volume} {100}},\ \bibinfo {pages} {165706} (\bibinfo {year}
  {2008})}\BibitemShut {NoStop}%
\bibitem [{\citenamefont {Ivanov}\ and\ \citenamefont
  {Abanov}(2013)}]{Ivanov2013b}%
  \BibitemOpen
  \bibfield  {author} {\bibinfo {author} {\bibfnamefont {D.~A.}\ \bibnamefont
  {Ivanov}}\ and\ \bibinfo {author} {\bibfnamefont {A.~G.}\ \bibnamefont
  {Abanov}},\ }\href {\doibase 10.1103/PhysRevE.87.022114} {\bibfield
  {journal} {\bibinfo  {journal} {Phys. Rev. E}\ }\textbf {\bibinfo {volume}
  {87}} (\bibinfo {year} {2013}),\ 10.1103/PhysRevE.87.022114},\ \Eprint
  {http://arxiv.org/abs/1203.6325} {arXiv:1203.6325} \BibitemShut {NoStop}%
\bibitem [{\citenamefont {Klich}\ and\ \citenamefont
  {Levitov}(2009)}]{klich_quantum_2009}%
  \BibitemOpen
  \bibfield  {author} {\bibinfo {author} {\bibfnamefont {I.}~\bibnamefont
  {Klich}}\ and\ \bibinfo {author} {\bibfnamefont {L.}~\bibnamefont
  {Levitov}},\ }\href {\doibase 10.1103/PhysRevLett.102.100502} {\bibfield
  {journal} {\bibinfo  {journal} {Physical Review Letters}\ }\textbf {\bibinfo
  {volume} {102}},\ \bibinfo {pages} {100502} (\bibinfo {year}
  {2009})}\BibitemShut {NoStop}%
\bibitem [{\citenamefont {Calabrese}\ \emph {et~al.}(2011)\citenamefont
  {Calabrese}, \citenamefont {Mintchev},\ and\ \citenamefont
  {Vicari}}]{Calabrese_2011}%
  \BibitemOpen
  \bibfield  {author} {\bibinfo {author} {\bibfnamefont {P.}~\bibnamefont
  {Calabrese}}, \bibinfo {author} {\bibfnamefont {M.}~\bibnamefont {Mintchev}},
  \ and\ \bibinfo {author} {\bibfnamefont {E.}~\bibnamefont {Vicari}},\ }\href
  {\doibase 10.1103/physrevlett.107.020601} {\bibfield  {journal} {\bibinfo
  {journal} {Physical Review Letters}\ }\textbf {\bibinfo {volume} {107}}
  (\bibinfo {year} {2011}),\ 10.1103/physrevlett.107.020601}\BibitemShut
  {NoStop}%
\bibitem [{\citenamefont {Vicari}(2012)}]{Vicarientangle}%
  \BibitemOpen
  \bibfield  {author} {\bibinfo {author} {\bibfnamefont {E.}~\bibnamefont
  {Vicari}},\ }\href {\doibase 10.1103/PhysRevA.85.062104} {\bibfield
  {journal} {\bibinfo  {journal} {Phys. Rev. A}\ }\textbf {\bibinfo {volume}
  {85}},\ \bibinfo {pages} {062104} (\bibinfo {year} {2012})}\BibitemShut
  {NoStop}%
\bibitem [{\citenamefont {Calabrese}\ \emph {et~al.}(2015)\citenamefont
  {Calabrese}, \citenamefont {Le~Doussal},\ and\ \citenamefont
  {Majumdar}}]{calabrese_random_2015}%
  \BibitemOpen
  \bibfield  {author} {\bibinfo {author} {\bibfnamefont {P.}~\bibnamefont
  {Calabrese}}, \bibinfo {author} {\bibfnamefont {P.}~\bibnamefont
  {Le~Doussal}}, \ and\ \bibinfo {author} {\bibfnamefont {S.~N.}\ \bibnamefont
  {Majumdar}},\ }\href {\doibase 10.1103/PhysRevA.91.012303} {\bibfield
  {journal} {\bibinfo  {journal} {Physical Review A}\ }\textbf {\bibinfo
  {volume} {91}},\ \bibinfo {pages} {012303} (\bibinfo {year}
  {2015})}\BibitemShut {NoStop}%
\bibitem [{\citenamefont {Garratt}\ \emph {et~al.}(2023)\citenamefont
  {Garratt}, \citenamefont {Weinstein},\ and\ \citenamefont
  {Altman}}]{garratt_measurements_2023}%
  \BibitemOpen
  \bibfield  {author} {\bibinfo {author} {\bibfnamefont {S.~J.}\ \bibnamefont
  {Garratt}}, \bibinfo {author} {\bibfnamefont {Z.}~\bibnamefont {Weinstein}},
  \ and\ \bibinfo {author} {\bibfnamefont {E.}~\bibnamefont {Altman}},\ }\href
  {\doibase 10.1103/PhysRevX.13.021026} {\bibfield  {journal} {\bibinfo
  {journal} {Physical Review X}\ }\textbf {\bibinfo {volume} {13}},\ \bibinfo
  {pages} {021026} (\bibinfo {year} {2023})}\BibitemShut {NoStop}%
\bibitem [{\citenamefont {Ivanov}\ \emph {et~al.}(2013)\citenamefont {Ivanov},
  \citenamefont {Abanov},\ and\ \citenamefont {Cheianov}}]{ivanov2013counting}%
  \BibitemOpen
  \bibfield  {author} {\bibinfo {author} {\bibfnamefont {D.~A.}\ \bibnamefont
  {Ivanov}}, \bibinfo {author} {\bibfnamefont {A.~G.}\ \bibnamefont {Abanov}},
  \ and\ \bibinfo {author} {\bibfnamefont {V.~V.}\ \bibnamefont {Cheianov}},\
  }\href {\doibase 10.1088/1751-8113/46/8/085003} {\bibfield  {journal}
  {\bibinfo  {journal} {J. Phys. A}\ }\textbf {\bibinfo {volume} {46}},\
  \bibinfo {pages} {085003} (\bibinfo {year} {2013})}\BibitemShut {NoStop}%
\bibitem [{\citenamefont {Fisher}(1984)}]{fisher_walks_1983}%
  \BibitemOpen
  \bibfield  {author} {\bibinfo {author} {\bibfnamefont {M.~E.}\ \bibnamefont
  {Fisher}},\ }\href {\doibase 10.1007/BF01009436} {\bibfield  {journal}
  {\bibinfo  {journal} {J. Stat. Phys}\ }\textbf {\bibinfo {volume} {34}},\
  \bibinfo {pages} {667} (\bibinfo {year} {1984})}\BibitemShut {NoStop}%
\bibitem [{\citenamefont {Kazakov}\ and\ \citenamefont
  {Migdal}(1988)}]{kazakov1988recent}%
  \BibitemOpen
  \bibfield  {author} {\bibinfo {author} {\bibfnamefont {V.}~\bibnamefont
  {Kazakov}}\ and\ \bibinfo {author} {\bibfnamefont {A.~A.}\ \bibnamefont
  {Migdal}},\ }\href {\doibase https://doi.org/10.1016/0550-3213(88)90146-0}
  {\bibfield  {journal} {\bibinfo  {journal} {Nuclear Physics B}\ }\textbf
  {\bibinfo {volume} {311}},\ \bibinfo {pages} {171} (\bibinfo {year}
  {1988})}\BibitemShut {NoStop}%
\bibitem [{\citenamefont {Majumdar}\ \emph {et~al.}(2009)\citenamefont
  {Majumdar}, \citenamefont {Nadal}, \citenamefont {Scardicchio},\ and\
  \citenamefont {Vivo}}]{majumdar_index_2009}%
  \BibitemOpen
  \bibfield  {author} {\bibinfo {author} {\bibfnamefont {S.~N.}\ \bibnamefont
  {Majumdar}}, \bibinfo {author} {\bibfnamefont {C.}~\bibnamefont {Nadal}},
  \bibinfo {author} {\bibfnamefont {A.}~\bibnamefont {Scardicchio}}, \ and\
  \bibinfo {author} {\bibfnamefont {P.}~\bibnamefont {Vivo}},\ }\href {\doibase
  10.1103/PhysRevLett.103.220603} {\bibfield  {journal} {\bibinfo  {journal}
  {Physical Review Letters}\ }\textbf {\bibinfo {volume} {103}},\ \bibinfo
  {pages} {220603} (\bibinfo {year} {2009})}\BibitemShut {NoStop}%
\bibitem [{\citenamefont {Marino}\ \emph
  {et~al.}(2014{\natexlab{a}})\citenamefont {Marino}, \citenamefont {Majumdar},
  \citenamefont {Schehr},\ and\ \citenamefont {Vivo}}]{Marino_vari}%
  \BibitemOpen
  \bibfield  {author} {\bibinfo {author} {\bibfnamefont {R.}~\bibnamefont
  {Marino}}, \bibinfo {author} {\bibfnamefont {S.~N.}\ \bibnamefont
  {Majumdar}}, \bibinfo {author} {\bibfnamefont {G.}~\bibnamefont {Schehr}}, \
  and\ \bibinfo {author} {\bibfnamefont {P.}~\bibnamefont {Vivo}},\ }\href
  {\doibase 10.1088/1751-8113/47/5/055001} {\bibfield  {journal} {\bibinfo
  {journal} {Journal of Physics A: Mathematical and Theoretical}\ }\textbf
  {\bibinfo {volume} {47}},\ \bibinfo {pages} {055001} (\bibinfo {year}
  {2014}{\natexlab{a}})}\BibitemShut {NoStop}%
\bibitem [{\citenamefont {Klich}(2002)}]{klich2002counting}%
  \BibitemOpen
  \bibfield  {author} {\bibinfo {author} {\bibfnamefont {I.}~\bibnamefont
  {Klich}},\ }\href@noop {} {\enquote {\bibinfo {title} {Full counting
  statistics: An elementary derivation of {L}evitov's formula},}\ } (\bibinfo
  {year} {2002}),\ \Eprint {http://arxiv.org/abs/cond-mat/0209642}
  {arXiv:cond-mat/0209642 [cond-mat.mes-hall]} \BibitemShut {NoStop}%
\bibitem [{\citenamefont {Deift}\ \emph {et~al.}(2011)\citenamefont {Deift},
  \citenamefont {Its},\ and\ \citenamefont
  {Krasovsky}}]{deift_asymptotics_2011}%
  \BibitemOpen
  \bibfield  {author} {\bibinfo {author} {\bibfnamefont {P.}~\bibnamefont
  {Deift}}, \bibinfo {author} {\bibfnamefont {A.}~\bibnamefont {Its}}, \ and\
  \bibinfo {author} {\bibfnamefont {I.}~\bibnamefont {Krasovsky}},\ }\href
  {\doibase 10.4007/annals.2011.174.2.12} {\bibfield  {journal} {\bibinfo
  {journal} {Annals of Mathematics}\ }\textbf {\bibinfo {volume} {174}},\
  \bibinfo {pages} {1243} (\bibinfo {year} {2011})}\BibitemShut {NoStop}%
\bibitem [{SM()}]{SM}%
  \BibitemOpen
  \href@noop {} {}\bibinfo {note} {See the Supplemental Material for deatils of
  calculations.}\BibitemShut {Stop}%
\bibitem [{\citenamefont {Cunden}\ \emph {et~al.}(2019)\citenamefont {Cunden},
  \citenamefont {Facchi}, \citenamefont {Ligab{\`o}},\ and\ \citenamefont
  {Vivo}}]{cunden2019third}%
  \BibitemOpen
  \bibfield  {author} {\bibinfo {author} {\bibfnamefont {F.~D.}\ \bibnamefont
  {Cunden}}, \bibinfo {author} {\bibfnamefont {P.}~\bibnamefont {Facchi}},
  \bibinfo {author} {\bibfnamefont {M.}~\bibnamefont {Ligab{\`o}}}, \ and\
  \bibinfo {author} {\bibfnamefont {P.}~\bibnamefont {Vivo}},\ }\href@noop {}
  {\bibfield  {journal} {\bibinfo  {journal} {Journal of statistical physics}\
  }\textbf {\bibinfo {volume} {175}},\ \bibinfo {pages} {1262} (\bibinfo {year}
  {2019})}\BibitemShut {NoStop}%
\bibitem [{\citenamefont {Marino}\ \emph
  {et~al.}(2014{\natexlab{b}})\citenamefont {Marino}, \citenamefont {Majumdar},
  \citenamefont {Schehr},\ and\ \citenamefont {Vivo}}]{Marino_2014}%
  \BibitemOpen
  \bibfield  {author} {\bibinfo {author} {\bibfnamefont {R.}~\bibnamefont
  {Marino}}, \bibinfo {author} {\bibfnamefont {S.~N.}\ \bibnamefont
  {Majumdar}}, \bibinfo {author} {\bibfnamefont {G.}~\bibnamefont {Schehr}}, \
  and\ \bibinfo {author} {\bibfnamefont {P.}~\bibnamefont {Vivo}},\ }\href
  {\doibase 10.1103/PhysRevLett.112.254101} {\bibfield  {journal} {\bibinfo
  {journal} {Phys. Rev. Lett.}\ }\textbf {\bibinfo {volume} {112}},\ \bibinfo
  {pages} {254101} (\bibinfo {year} {2014}{\natexlab{b}})}\BibitemShut
  {NoStop}%
\bibitem [{\citenamefont {Marino}\ \emph {et~al.}(2016)\citenamefont {Marino},
  \citenamefont {Majumdar}, \citenamefont {Schehr},\ and\ \citenamefont
  {Vivo}}]{marino_number_2016}%
  \BibitemOpen
  \bibfield  {author} {\bibinfo {author} {\bibfnamefont {R.}~\bibnamefont
  {Marino}}, \bibinfo {author} {\bibfnamefont {S.~N.}\ \bibnamefont
  {Majumdar}}, \bibinfo {author} {\bibfnamefont {G.}~\bibnamefont {Schehr}}, \
  and\ \bibinfo {author} {\bibfnamefont {P.}~\bibnamefont {Vivo}},\ }\href
  {\doibase 10.1103/PhysRevE.94.032115} {\bibfield  {journal} {\bibinfo
  {journal} {Physical Review E}\ }\textbf {\bibinfo {volume} {94}},\ \bibinfo
  {pages} {032115} (\bibinfo {year} {2016})}\BibitemShut {NoStop}%
\bibitem [{\citenamefont {Smith}\ \emph {et~al.}(2021)\citenamefont {Smith},
  \citenamefont {Le~Doussal}, \citenamefont {Majumdar},\ and\ \citenamefont
  {Schehr}}]{smith_full_2021}%
  \BibitemOpen
  \bibfield  {author} {\bibinfo {author} {\bibfnamefont {N.}~\bibnamefont
  {Smith}}, \bibinfo {author} {\bibfnamefont {P.}~\bibnamefont {Le~Doussal}},
  \bibinfo {author} {\bibfnamefont {S.}~\bibnamefont {Majumdar}}, \ and\
  \bibinfo {author} {\bibfnamefont {G.}~\bibnamefont {Schehr}},\ }\href
  {\doibase 10.21468/SciPostPhys.11.6.110} {\bibfield  {journal} {\bibinfo
  {journal} {SciPost Physics}\ }\textbf {\bibinfo {volume} {11}},\ \bibinfo
  {pages} {110} (\bibinfo {year} {2021})}\BibitemShut {NoStop}%
\bibitem [{\citenamefont {Bocini}\ and\ \citenamefont
  {St\'ephan}(2021)}]{Bocini_2021}%
  \BibitemOpen
  \bibfield  {author} {\bibinfo {author} {\bibfnamefont {S.}~\bibnamefont
  {Bocini}}\ and\ \bibinfo {author} {\bibfnamefont {J.-M.}\ \bibnamefont
  {St\'ephan}},\ }\href {\doibase 10.1088/1742-5468/abcd34} {\bibfield
  {journal} {\bibinfo  {journal} {Journal of Statistical Mechanics: Theory and
  Experiment}\ }\textbf {\bibinfo {volume} {2021}},\ \bibinfo {pages} {013204}
  (\bibinfo {year} {2021})}\BibitemShut {NoStop}%
\end{thebibliography}%


\begin{thebibliography}{9}%
\makeatletter
\providecommand \@ifxundefined [1]{%
 \@ifx{#1\undefined}
}%
\providecommand \@ifnum [1]{%
 \ifnum #1\expandafter \@firstoftwo
 \else \expandafter \@secondoftwo
 \fi
}%
\providecommand \@ifx [1]{%
 \ifx #1\expandafter \@firstoftwo
 \else \expandafter \@secondoftwo
 \fi
}%
\providecommand \natexlab [1]{#1}%
\providecommand \enquote  [1]{``#1''}%
\providecommand \bibnamefont  [1]{#1}%
\providecommand \bibfnamefont [1]{#1}%
\providecommand \citenamefont [1]{#1}%
\providecommand \href@noop [0]{\@secondoftwo}%
\providecommand \href [0]{\begingroup \@sanitize@url \@href}%
\providecommand \@href[1]{\@@startlink{#1}\@@href}%
\providecommand \@@href[1]{\endgroup#1\@@endlink}%
\providecommand \@sanitize@url [0]{\catcode `\\12\catcode `\$12\catcode
  `\&12\catcode `\#12\catcode `\^12\catcode `\_12\catcode `\%12\relax}%
\providecommand \@@startlink[1]{}%
\providecommand \@@endlink[0]{}%
\providecommand \url  [0]{\begingroup\@sanitize@url \@url }%
\providecommand \@url [1]{\endgroup\@href {#1}{\urlprefix }}%
\providecommand \urlprefix  [0]{URL }%
\providecommand \Eprint [0]{\href }%
\providecommand \doibase [0]{http://dx.doi.org/}%
\providecommand \selectlanguage [0]{\@gobble}%
\providecommand \bibinfo  [0]{\@secondoftwo}%
\providecommand \bibfield  [0]{\@secondoftwo}%
\providecommand \translation [1]{[#1]}%
\providecommand \BibitemOpen [0]{}%
\providecommand \bibitemStop [0]{}%
\providecommand \bibitemNoStop [0]{.\EOS\space}%
\providecommand \EOS [0]{\spacefactor3000\relax}%
\providecommand \BibitemShut  [1]{\csname bibitem#1\endcsname}%
\let\auto@bib@innerbib\@empty
\bibitem [{\citenamefont {Szeg\H{o}}(1939)}]{szeg1939orthogonal}%
  \BibitemOpen
  \bibfield  {author} {\bibinfo {author} {\bibfnamefont {G.}~\bibnamefont
  {Szeg\H{o}}},\ }\href@noop {} {\emph {\bibinfo {title} {Orthogonal
  polynomials}}},\ Vol.~\bibinfo {volume} {23}\ (\bibinfo  {publisher}
  {American Mathematical Soc.},\ \bibinfo {year} {1939})\BibitemShut {NoStop}%
\bibitem [{\citenamefont {Johanson}(1988)}]{Johansson_1988}%
  \BibitemOpen
  \bibfield  {author} {\bibinfo {author} {\bibfnamefont {K.}~\bibnamefont
  {Johanson}},\ }\href@noop {} {\bibfield  {journal} {\bibinfo  {journal}
  {Bull. Sc. math. 2me s\'erie}\ }\textbf {\bibinfo {volume} {112}},\ \bibinfo
  {pages} {257} (\bibinfo {year} {1988})}\BibitemShut {NoStop}%
\bibitem [{\citenamefont {Bump}\ and\ \citenamefont
  {Diaconis}(2002)}]{bump2002toeplitz}%
  \BibitemOpen
  \bibfield  {author} {\bibinfo {author} {\bibfnamefont {D.}~\bibnamefont
  {Bump}}\ and\ \bibinfo {author} {\bibfnamefont {P.}~\bibnamefont
  {Diaconis}},\ }\href@noop {} {\bibfield  {journal} {\bibinfo  {journal}
  {Journal of Combinatorial Theory, Series A}\ }\textbf {\bibinfo {volume}
  {97}},\ \bibinfo {pages} {252} (\bibinfo {year} {2002})}\BibitemShut
  {NoStop}%
\bibitem [{\citenamefont {Garc{\'\i}a-Garc{\'\i}a}\ and\ \citenamefont
  {Tierz}(2020)}]{garcia2020matrix}%
  \BibitemOpen
  \bibfield  {author} {\bibinfo {author} {\bibfnamefont {D.}~\bibnamefont
  {Garc{\'\i}a-Garc{\'\i}a}}\ and\ \bibinfo {author} {\bibfnamefont
  {M.}~\bibnamefont {Tierz}},\ }\href@noop {} {\bibfield  {journal} {\bibinfo
  {journal} {Journal of Physics A: Mathematical and Theoretical}\ }\textbf
  {\bibinfo {volume} {53}},\ \bibinfo {pages} {345201} (\bibinfo {year}
  {2020})}\BibitemShut {NoStop}%
\bibitem [{Note1()}]{Note1}%
  \BibitemOpen
  \bibinfo {note} {The formula can be easily obtained from the linearity of the
  determinant of (\ref {Tsymbol-100}) with respect to columns. For details see
  \cite {andreief1886note,forrester2019meet}.}\BibitemShut {Stop}%
\bibitem [{\citenamefont {Deift}\ \emph {et~al.}(2011)\citenamefont {Deift},
  \citenamefont {Its},\ and\ \citenamefont
  {Krasovsky}}]{deift_asymptotics_2011}%
  \BibitemOpen
  \bibfield  {author} {\bibinfo {author} {\bibfnamefont {P.}~\bibnamefont
  {Deift}}, \bibinfo {author} {\bibfnamefont {A.}~\bibnamefont {Its}}, \ and\
  \bibinfo {author} {\bibfnamefont {I.}~\bibnamefont {Krasovsky}},\ }\href
  {\doibase 10.4007/annals.2011.174.2.12} {\bibfield  {journal} {\bibinfo
  {journal} {Annals of Mathematics}\ }\textbf {\bibinfo {volume} {174}},\
  \bibinfo {pages} {1243} (\bibinfo {year} {2011})}\BibitemShut {NoStop}%
\bibitem [{Note2()}]{Note2}%
  \BibitemOpen
  \bibinfo {note} {The factor 1/2 comes from the normalisation: the density in
  momentum space is $\sigma (k)/2\pi $ and $G'(k) =\pi \sigma (k) /2\pi =
  \sigma (k)/2$.}\BibitemShut {Stop}%
\bibitem [{\citenamefont {Andr{\'e}ief}(1886)}]{andreief1886note}%
  \BibitemOpen
  \bibfield  {author} {\bibinfo {author} {\bibfnamefont {C.}~\bibnamefont
  {Andr{\'e}ief}},\ }\href@noop {} {\bibfield  {journal} {\bibinfo  {journal}
  {M{\'e}m. de la Soc. Sci. Bordeaux}\ }\textbf {\bibinfo {volume} {2}},\
  \bibinfo {pages} {1} (\bibinfo {year} {1886})}\BibitemShut {NoStop}%
\bibitem [{\citenamefont {Forrester}(2019)}]{forrester2019meet}%
  \BibitemOpen
  \bibfield  {author} {\bibinfo {author} {\bibfnamefont {P.~J.}\ \bibnamefont
  {Forrester}},\ }\href@noop {} {\bibfield  {journal} {\bibinfo  {journal}
  {Random Matrices: Theory and Applications}\ }\textbf {\bibinfo {volume}
  {8}},\ \bibinfo {pages} {1930001} (\bibinfo {year} {2019})}\BibitemShut
  {NoStop}%
\end{thebibliography}%

\end{document}


\title{
Supplemental material: 
Phase transitions in full counting statistics of free fermions and directed polymers}
\author{James S. Pallister}
\affiliation{School of Physics and Astronomy, University of Birmingham, Edgbaston, Birmingham, B15 2TT, UK }

\author{Samuel H. Pickering}
\affiliation{School of Physics and Astronomy, University of Birmingham, Edgbaston, Birmingham, B15 2TT, UK }

\author{Dimitri M. Gangardt}
\affiliation{School of Physics and Astronomy, University of Birmingham, Edgbaston, Birmingham, B15 2TT, UK }

\author{Alexander G. Abanov }
\affiliation{Department of Physics and Astronomy, Stony Brook University, 
Stony Brook, NY 11794, USA
}

\date{\today}

\maketitle

\renewcommand{\theequation}{S.\arabic{equation}}

\section{Toeplitz determinant as a Coulomb gas}
\label{sec:ToeplitzCoulomb}
The equivalence between the finite-size Toeplitz determinant and the Coulomb gas partition function is well known. It appeared already in works of Heine and Szeg\H{o} (see for example, the formula (2.2.11) in Ref.~\cite{szeg1939orthogonal}). In Ref.~\cite{Johansson_1988} it is referred to as a "well-known integral representation of a Toeplitz determinant".
Written as an integral over unitary group it is sometimes referred to as the Heine-Szeg\H{o} formula, see section 2 of Ref.~\cite{bump2002toeplitz}. In the latter reference the formula was generalized to Toeplitz minors. For a recent work on further generalizations including other groups, see Ref.~\cite{garcia2020matrix}. To make the paper more self-contained we present here the derivation of that equivalence. 

Let us consider the Toeplitz determinant $\mathrm{det}(T)$ of the $R\times R$ matrix defined by the Toeplitz symbol $t(k)$ as 
\begin{align}
    T_{mn} = \mathrm \int \frac{dk}{2\pi} t(k) e^{\ii k(m-n) }\, ,
 \label{Tsymbol-100}
\end{align}
and $n,m=1,\ldots R$. Using Andr\'eief's integration formula \footnote{The formula can be easily obtained from the linearity of the determinant of (\ref{Tsymbol-100}) with respect to columns. For details see \cite{andreief1886note,forrester2019meet}. }, we obtain
\begin{align}
    \mathrm{det}(T) &=\frac{1}{R!}\int \frac{dk_1}{2\pi}\ldots\int\frac{dk_R}{2\pi} \,
 \mathrm{det}\Big( t(k_j)e^{\ii k_j (m-1)}\Big)_{j,m=1}^R \mathrm{det} \Big(e^{-\ii k_j (n-1)}\Big)_{j,n=1}^R,
 \nonumber
\end{align}
and then using Vandermonde formula 
\begin{align}
    \mathrm{det}(T) &= \frac{1}{R!}\int \frac{d^Rk}{(2\pi)^R} \left|\Delta(e^{ik})\right|^2 \prod_{j=1}^R t(k_j)\,,
 \label{toeplitz-500}
\end{align}
where the Vandermonde determinant is defined as
\begin{align}
    \Delta(e^{ik}) \equiv \prod_{1\leq n<m\leq R} \left(e^{\ii k_m}-e^{\ii k_n}\right)\,.
\end{align}
Exponentiating the integrand in (\ref{toeplitz-500}) we obtain a partition function for the Coulomb gas
\begin{align}
    \mathrm{det}(T) &= \frac{1}{R!}\int \frac{d^Rk}{(2\pi)^R} e^{-F\left(e^{\ii k}\right)}\,,
 \label{toeplitz-510} \\
    F\left(e^{\ii k}\right) &= -\sum_{1\leq n<m\leq R}\log\left|e^{\ii k_n}-e^{\ii k_m}\right|^2 -\sum_{n=1}^R \log\Big(t(k_n)\Big)\,.
 \nonumber
\end{align}
In this representation, the logarithm of the Toeplitz symbol enters as an external potential applied to the Coulomb gas on a unit circle. 

We will apply the general Coulomb-gas representation of the Toeplitz determinant (\ref{toeplitz-510}) to the FCS generating function.

\section{  Toeplitz determinant representation of FCS and solution of the corresponding  Coulomb gas problem }
\label{sec:CoulombFCS}
For FCS generating function we get the Toeplitz determinant  and the corresponding Coulomb gas partition function with  the symbol
\begin{align}
    t(k) =\frac{ 1+e^{-s} e^{-\beta h(k)}}{1+e^{-\beta h(k)}} = e^{-s v(k)} \, , 
\end{align}
For finite $\beta$ this symbol is smooth, but  
at zero temperature, $1/\beta=0$ it develops  step-like singularities, resulting in $v(k) = 1$ in the interval $-k_F<k<k_F$ and $0$ otherwise, that is $v(k)=\theta(k_F-|k|)$. 
Asymptotics  of the corresponding Toeplitz determinant can be obtained by standard analysis   \cite{deift_asymptotics_2011} via Szeg\H{o} theorem and Hartwig-Fischer conjecture. The situation becomes more complicated in the scaling limit $s=\alpha R$. This is where the Coulomb gas representation 
\begin{align}\label{eq:chiferm1}
    \chi_R(s) = \frac{1}{R!}
    \int  \frac{d^R k}{(2\pi)^R}  \left|\Delta(e^{\ii k
})\right|^2 \prod_{l=1}^R e^{-s v(k_l)} ,
\end{align}
becomes indispensable. 
This integral \eqref{eq:chiferm1} is dominated by stationary configurations of charges. To find it we assume charges are smoothly distributed on support $\gamma$ with  density    
\begin{align}\label{eq:norm}
\sigma(k)
 = \frac{2\pi}{R} \sum_i \delta(k-k_i)\, ,\qquad
\int_\gamma \frac{\dd k}{2\pi} \sigma (k) =1 \, .    
\end{align}
and represent Eq.~\eqref{eq:chiferm1} as the path integral
\begin{align}
 \label{eq:S1}
    \chi_R (\alpha R) =e^{-R^2f(\alpha)} 
    =    \int\mathrm{D}[\sigma] \dd\mu\,
    e^{-R^2 S[\sigma] }\,,
\end{align}
over configurations $\sigma(k)$ weighted by the action
\begin{align}
 \label{eq:z1}
    S[\sigma] 
    &=-\frac{1}{2} \int_{\gamma}\dk\int_{\gamma}\dk' \sigma(k)\log\left|e^{\ii k}-e^{\ii k'}\right|^2 \sigma(k')   +\alpha \int_{\gamma}\frac{\dd k}{2\pi} \sigma(k) v(k) -\mu\left(\int_{\gamma}\dk \sigma (k) - 1\right)  \, .
\end{align} 
The chemical potential $\mu$ acts as a Lagrange multiplier to ensure the normalisation in  Eq.~\eqref{eq:norm}. The measure of integration $\mathrm{D}[\sigma]$ is chosen so that  $\chi_R(0)=1$, or, equivalently, $f(0)=0$. Moving from the microscopic momenta $\{k_i\}$ to the macroscopic density $\sigma$ generates $\mathcal{O}(R)$ entropic terms in the exponent. These are subdominant in our large deviation scaling limit and thus neglected. 

Taking variation of the action w.r.t. $\sigma(k) $  we arrive to the saddle-point equation
\begin{align}
\label{eq:mueq}
  \mu =\alpha v(k)  -\int_{\gamma}  \log\abs{e^{\ii k}-e^{\ii k'}}^2\sigma(k')\frac{\dd k'}{2\pi} .
\end{align}
valid in the support $k\in\gamma$.

\begin{figure}[t]
\centering
  \includegraphics[width=0.44\textwidth]
   {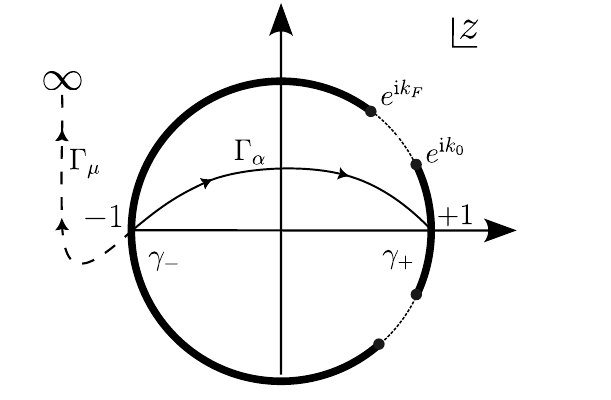}
  \includegraphics[width=0.44\textwidth]
  {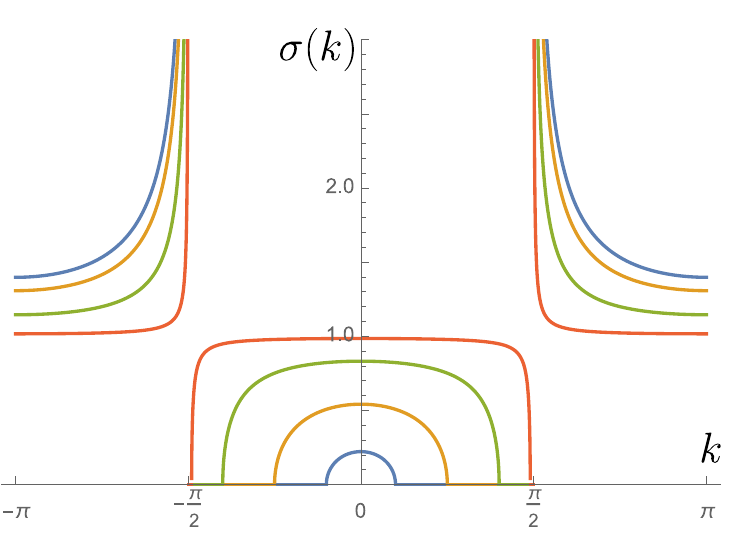}
    \caption{Left: Complex plane structure for Coulomb gas problem, $z = e^{\ii k}$. The charges reside in two cuts specified by $z_\mathrm{F}$ and $z_\alpha$. Complete emptiness corresponds to $z_\alpha = 1$. Contours of integration in Eqs.~\eqref{eq:alpha},\eqref{eq:mu} are shown. Right: density of charges $\sigma(k)$, Eq.~\eqref{eq:sigmak} for $\nu_0 =k_0/\pi = 0.1,0.25,0.4,0.49$. }
    \label{fig:contour}
\end{figure}

We introduce the complex potential
\begin{align}
    \varphi(z) & = \int_{\gamma}\frac{\dd k}{2\pi}\log(z-e^{\ii k})\sigma(k)\, ,
\end{align}
analytic in the whole complex plane of $z$ apart from the support $z=e^{\ii k}$, $k\in \gamma$.  As a result of the discontinuity of the potential $v(k)$ we assume the support consists of two disjoint intervals $\gamma=\gamma_+\cup \gamma_-$. This is illustrated in the left panel of Fig.~\ref{fig:contour}.
Choosing two points $k_-=\pi,k_+=0$ 
on the support $\gamma$ such that  $v(k_-)=0$ and $v(k_+)=1$,   
Eq.~\eqref{eq:mueq} becomes equivalent to
\begin{align}\label{eq:erdung0}
    2\re\varphi(-1) &= -\mu\,, 
 \\ \label{eq:erdung1}
    2\re\varphi(+1) &=\alpha -\mu\,.
\end{align}
Additionally, as a consequence of the normalisation condition, Eq.~\eqref{eq:norm},  $\varphi(z)=\log z$ for $z\to\infty$. This condition together with  Eqs.~\eqref{eq:erdung0},\eqref{eq:erdung1} 
fully determine the optimal $\varphi(z)$. 

The support $\gamma$ and the density $\sigma(k)$ are  obtained from the non-analyticity of the resolvent $g(z) +1/2 = z\der_z\varphi(z)$. 
The density of charges is given by the jump across the cuts,  $ \sigma (k)=
g_+(e^{\ii k}) - g_-(e^{\ii k})  = 2 \re g(e^{\ii k}) $.
Taking the derivative of both sides of Eq.~\eqref{eq:mueq}   we obtain the condition
\begin{align}\label{eq:em1}
g_+(e^{\ii k})+g_-(e^{\ii k}) = 0\, , \quad k\in \gamma\, ,\, k\neq\pm k_F\, .
\end{align}
Allowing for  possible empty intervals, $\sigma(k)=0$, leads to  $g_+^2(z)=g_-^2(z)$, so $g^2(z)$ is analytic
in the whole complex plane apart from  poles at $z=e^{\pm\ii k_F}$ and $z=\infty$, where it behaves as $g^2(z)=1/4+1/z+\mathcal{O}(z^{-2})$ due to normalisation.  This fixes completely the form of the resolvent,
\begin{align}
    g^2(z) = \frac{1}{4} +\frac{zd}{(z-e^{\ii k_F})(z-e^{-\ii k_F})}\, .
\end{align}
The parameter $d$ is written in the form $d=\sin^2\frac{k_0}{2}-\sin^2\frac{k_F}{2}$ so that
\begin{align}
    g(z) = \frac{1}{2}\sqrt{
    \frac{(z-e^{\ii k_0})(z-e^{-\ii k_0})}{(z-e^{\ii k_F})(z-e^{-\ii k_F})}
    } ,
 \label{eq:gz-200}
\end{align}
coincides with 
Eq.~(4) in the main text. The corresponding density of charges is
\begin{align}\label{eq:sigmak}
    \sigma(k) = \sqrt{\frac{\sin^2 \frac{k}{2}-\sin^2\frac{k_0}{2}}{\sin^2 \frac{k}{2}-\sin^2\frac{k_F}{2}}} \, ,
\end{align}
and is shown in the right panel of Fig.~\ref{fig:contour} for several values of $k_0=\pi\nu_0$.

Using Eqs.~\eqref{eq:erdung1} and \eqref{eq:erdung0}  and $\p_z\phi= (g(z)+1/2)/z$ we obtain
\begin{align}\label{eq:alpha}
    \alpha &=2\re \left(\varphi(+1)-\varphi(-1)\right)=2 \re\int_{\Gamma_\alpha} \frac{g(z)\dd z}{z}  =2 \re\int_{-1}^{+1} \frac{g(z)\dd z}{z}\,,  
 \\ \label{eq:mu1}
    \mu &=  -2\re \varphi(-1)  =2 \re\int_{\Gamma_\mu}\frac{g(z)\dd z}{z}=2 \re\int_{-1}^{\infty} \frac{g(z)\dd z}{z} \,. 
\end{align}
The integration countours $\Gamma_{\alpha,\mu}$ are shown in the left panel of Fig.\ref{fig:contour}. They can be deformed as long as their end-points remain on the same intervals $\gamma_\pm$ of the support $\gamma$. The integral for $\mu$ diverges logarithmically at infinity and must be regularised by subtracting $\mu_0 = \int_{-1}^\infty \dd z/z$  corresponding to $k_0=k_F$ and $\alpha=0$.

Taking variation of the action, Eq.~\eqref{eq:z1}, w.r.t. $\alpha$ and using  the saddle-point equation, Eq.~\eqref{eq:mueq} we get
\begin{align}\label{eq:nuI}
    f'(\alpha) =\frac{\der S}{\der\alpha} = \int_{-k_F}^{k_F} \frac{\dd k }{2\pi} \sigma(k) =\nu_I,
\end{align}
which is equivalent to 
Eq.~(6) in the main text as for $z=e^{\ii k}$ we have $\sigma(k)\dd k  = 2\Im g (z)\dd z/z $.

The phase transition happens when  only  one  branch cut is left for $\nu_0=0$. The integral in \niceref{eq:alpha} is elementary for $g(z)$ in (\ref{eq:gz-200}) at  $k_0=0$ providing the critical value, 
Eq.~(7) in the main text.

Similarly, using the normalisation \niceref{eq:norm} and  the saddle-point equations Eq.~\eqref{eq:mueq}, 
to eliminate the dependence on the log term in \niceref{eq:z1} in favour of $\alpha, \mu$ we get 
\begin{align}
f(\alpha)=\frac{\mu}{2} + \frac{\alpha\nu_I}{2} \, .
\end{align}
yielding  the correct normalisation $f(0)=0$. For $\alpha\ge\alpha_\mathrm{c}$ we have $\nu_I=0$
and $f=\mu/2$. For this regime the chemical 
potential, Eq.~\eqref{eq:mu1}, becomes an elementary integral and we find $ \mu = -\log\cos^2\frac{k_F}{2}$, leading to 
Eq.~(8) in the main text.

\section{Evaluation of integrals in the critical regime}

We obtain the leading behaviour of $\alpha$, Eq.~\eqref{eq:alpha}, in the limit $k_0
\to 0$.  Defining $a=\sin\frac{k_0}{2}$, $b=\sin\frac{k_F}{2} $ and  
using $z=\mp e^{-v}$ in the negative/positive  part  of the interval $[-1,1]=[-1,0]\cup [0,1]$, Eq.~\eqref{eq:alpha} becomes the sum of two contributions $\alpha=I\left(\sqrt{1-a^2},\sqrt{1-b^2}\right)-I(a,b)$, where subtracting one for convergence
\begin{align}
    I(a,b)=\int_0^\infty \dd v \left(\sqrt{\frac{\sinh^2\frac{v}{2}+a^2}{\sinh^2\frac{v}{2}+b^2}}-1\right)\,
\end{align}

The critical value $\alpha_\mathrm{c}$ corresponds to $a=0$, which gives
\begin{align}
    \alpha_c =I\left(1,\sqrt{1-b^2}\right) -I(0,b) &=\int_0^\infty \dd v\left(\frac{\cosh\frac{v}{2}}{\sqrt{\cosh^2\frac{v}{2}-b^2}}-\frac{\sinh\frac{v}{2}}{\sqrt{\sinh^2\frac{v}{2}+b^2}}\right)\notag \\&= 2\arcosh\frac{1}{\sqrt{1-b^2}}\, ,  
\end{align}
identical to 
Eq.~(7) in the main text. We proceed with calculating $\alpha-\alpha_\mathrm{c} $ by combining the terms
\begin{align}
    \alpha-\alpha_\mathrm{c} &= \left[I\left(\sqrt{1-a^2},\sqrt{1-b^2}\right)-I\left(1,\sqrt{1-b^2}\right) \right]\notag \\&-\left[I\left(a,b\right)-I\left(0,b\right) \right]=I_1-I_2.
\end{align}
The first term can be approximated to the leading order,
\begin{align}\label{eq:i1}
    I_1&=I\left(\sqrt{1-a^2},\sqrt{1-b^2}\right)-I\left(1,\sqrt{1-b^2}\right) = \int_0^\infty \dd v \frac{\sqrt{\cosh^2\frac{v}{2}-a^2}-\cosh\frac{v}{2}}{\sqrt{\sinh^2\frac{v}{2}+1-b^2}}\notag \\&\simeq-\frac{a^2}{2} \int_0^\infty \frac{\dd v }{\cosh\frac{v}{2}\sqrt{\sinh^2\frac{v}{2}+1-b^2}} = -a^2 \int_0^\infty \frac{\dd u }{\cosh^2 u-b^2\sinh^2u} = -\frac{a^2}{b}\artanh{b}\, , 
\end{align}
with the change  of variables $\sinh\frac{v}{2}=\sqrt{1-b^2}\sinh u$, $\cosh\frac{v}{2} \dd v = 2\sqrt{1-b^2}\cosh u\,\dd u$.

To evaluate the second term we divide the integration domain into two intervals,
\begin{align}
    I_2 &=I\left(a,b\right)-I\left(0,b\right) = \left(\int_0^\lambda+\int_\lambda^\infty\right) \dd v \frac{\sqrt{\sinh^2\frac{v}{2}+a^2}-\sinh\frac{v}{2}}{\sqrt{\sinh^2\frac{v}{2}+b^2}}\notag \\ &\simeq \frac{1}{b}\int_0^\lambda \dd v \left(\sqrt{\left(\frac{v}{2}\right)^2+a^2}-\frac{v}{2}\right) + \frac{a^2}{2}\int_\lambda^\infty  \frac{\dd v}{\sinh\frac{v}{2}\sqrt{\sinh^2\frac{v}{2}+b^2}},
\end{align}
with the corresponding approximations valid provided we chose $a\ll\lambda\ll b$. The first integral is elementary and after change of variables $v=2a x$ becomes
\begin{align}
    &\frac{2 a^2}{b}\left[\int_0^{\lambda/2a} \dd x \sqrt{x^2+1}-\frac{1}{2}\left(\frac{\lambda}{2a}\right)^2\right] = \frac{a^2}{b}\left[\frac{\lambda}{2a}\sqrt{1+\left(\frac{\lambda}{2a}\right)^2}- \left(\frac{\lambda}{2a}\right)^2+\arsinh\frac{\lambda}{2a}\right]\notag \\&\simeq 
    \frac{a^2}{b}\left[\frac{1}{2}+\log\frac{\lambda}{a} \right]\, .
\end{align}
The second integral is evaluated similarly to Eq.~\eqref{eq:i1} by changing variable $\cosh\frac{v}{2}=\sqrt{1-b^2}\cosh u$,
\begin{align}\label{eq:i22}
    &\frac{a^2}{2} \int_\lambda^\infty \frac{\dd v }{\sinh\frac{v}{2}\sqrt{\sinh^2\frac{v}{2}+b^2}} =- \frac{a^2}{b^2} \int_{\lambda'}^\infty \frac{\dd u }{\cosh^2 u-b^{-2}\sinh^2u} \notag \\
&= -\frac{a^2}{b}\left(\artanh{b}-\artanh{\frac{b}{\tanh\lambda'}}\right)\, .
\end{align}

Here $\cosh\lambda' = \cosh (\lambda/2) /\sqrt{1-b^2}$, so that $1-b/\tanh{\lambda'} \simeq (1-b^2)\lambda^2/8 b^2$ and $2\artanh(b/\tanh\lambda') \simeq \log ((16/\lambda^2) b^2/(1-b^2))$.
Using this we get the answer
\begin{align}
   \alpha-\alpha_\mathrm{c} &\simeq -\frac{a^2}{2b} \left[1+2\log\frac{\lambda}{a
} -\log \left(\frac{1-b^2}{b^2}\frac{\lambda^2}{16}\right)\right]\notag\\
&=-\frac{a^2}{2b} \left[1+\log \left(\frac{16}{a^2}\frac{b^2}{1-b^2}\right)\right],
\end{align}
independent of the cut-off $\lambda$. 

To calculate the chemical potential
we use Eq.~\eqref{eq:erdung0}.
Subtracting the result for $\alpha=0$ containing the same logarithmic divergence, we have
\begin{align}\label{eq:mu}
    \mu -\mu_0=2 \re \int_{e^{\ii k_0}}^\infty \left(\frac{g(z)}{z}-\frac{1}{2z}\right) \dd z\, .
\end{align}
Taking $k_0=\pi$ and using $z=-e^v$ this integral becomes
\begin{align}\label{eq:mu100}
    \mu -\mu_0= I\left(\sqrt{1-a^2},\sqrt{1-b^2}\right) 
    - I (1,1)\, .
\end{align}
For $a=0$ the integral is elementary
\begin{align}
   \mu-\mu_0 &= \int_0^\infty \left(\frac{\cosh\frac{v}{2}\dd v}{\sqrt{\sinh^2\frac{v}{2}+1-b^2}} -1\right) =  \log\frac{1}{1-b^2}.
\end{align}

Using the resolvent, 
Eq.~(4), the integrals in Eqs.~\eqref{eq:alpha}, \eqref{eq:nuI} can be evaluated for general values of $a,b$ as well in terms of  the incomplete and complete  elliptic integral of the third kind $\Pi(n;\phi|m)$ and $\Pi(n|m)=\Pi\left(n;\frac{\pi}{2}|m\right)$. We obtain
\begin{align}
\label{eq:fillinga}
    \alpha & =  -\frac{2 a^2}{b\sqrt{1-a^2}}\Pi\left(\left.\frac{1}{1-a^2}\right\vert \frac{1-\frac{a^2}{b^2}}{1-a^2} \right)\,, 
 \\
    1-\nu_I &=\int_{k_F}^{2\pi - k_F} \frac{\dd k}{2\pi} \sigma(k)
    = \frac{2}{\pi} \frac{(1-a^2)}{\sqrt{b^2-a^2}}\Pi\left(\left. \frac{1-b^2}{a^2-b^2} \right\vert \frac{1-b^2}{1-\frac{b^2}{a^2}} \right)\,,
 \notag \\
    \mu(a,b) &=\frac{2 (1-a^2)}{a \sqrt{1-b^2}}\Pi\left(\left.\frac{1}{a^2}; \arcsin{a}\right\vert \frac{1-\frac{b^2}{a^2}}{1-b^2} \right),
\end{align}
where $a = \sin{\frac{k_0}{2}}, b = \sin{\frac{k_F}{2}}$.

\section{Hydrodynamics and Coulomb Gas solution}

It is useful to paramterize the hydrodynamical solution via the complex potential $\chi(x,\tau)$ such that  
\begin{align}
    \der_x \chi(x,\tau) = k(x,\tau) = \pi \nu (x,\tau) +\ii \upsilon (x,\tau)\, .
 \label{derchi}
\end{align}
and represent it via Legendre transform 
\begin{align}
    \chi(x,\tau)  & = kx -G(k) + \ii \varepsilon (k) \tau 
 \label{chixtau}\\
    x +\ii \varepsilon'(k) \tau &= G'(k)
 \label{xtauGprime}
\end{align}
This way the hydrodynamic equations of motion Eq. (20) is automatically satisfied. Note that the time $\tau$ is treated as an external parameter.

\begin{figure}[t]
\centering
  \includegraphics[width=0.47\textwidth]
   {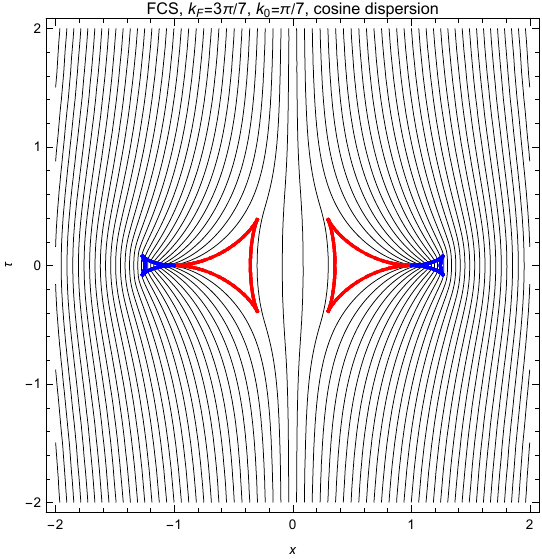}
  \includegraphics[width=0.47\textwidth]
  {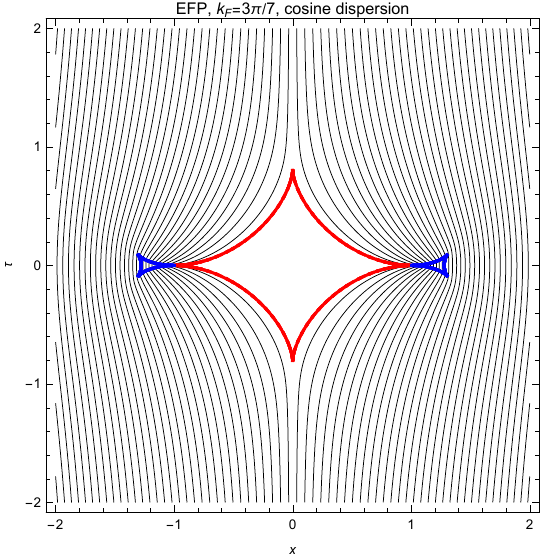}
    \caption{Left: Polymer configuration is shown for $k_F=3\pi/7$, $k_0=\pi/7$. the boundaries of completely empty regions ($\nu=0$) and regions with maximal density ($\nu=1$) are shown in red and blue, respectively.   Right: Polymer configuration is shown for $k_F=3\pi/7$ and $\alpha>\alpha_c$. }
    \label{fig:trajectories}
\end{figure}

From (\ref{derchi}) it follows that the density of polymers is given by
\begin{align}
    \nu(x,\tau) = \frac{1}{\pi}\Re \p_x \chi(x,\tau)
\end{align}
and, therefore, one can write the following function
\begin{align}
    n(x,\tau) = \int_0^x\nu(x,\tau)\,dx=\frac{1}{\pi}\Re \Big(\chi(x,\tau)-\chi(0,\tau)\Big)
\end{align}
as a counting function of polymers so that the solution $x_l(\tau)$ of the equation $n(x,\tau)=l$ gives the trajectory of the polymer number $l$. We show some of those trajectories in Fig.~\ref{fig:trajectories}.

Focusing on initial state at $\tau=0$  we have
$kx = \chi(x) +G(k)$. Interpreting the left hand side of this equation as phase-space volume occupied by $n$ particles with $k=k_n$ and $x=x_n$. Increasing the number of particles by one, $n\to n+1 = n+\dd n$ we compute
\begin{align}
    2\pi \dd n =  \dd (k x) = \dd \chi (x) +\dd G(k) = \pi \nu (x) \dd x + G'(k) \dd k\, .
\end{align}
As $\nu (x) = \dd n/\dd x$ we must have $G'(k) = \dd n /\dd k= \sigma (k)/2$ \footnote{The factor 1/2 comes from the normalisation:  the density in  momentum space is  $\sigma(k)/2\pi $ and $G'(k) =\pi \sigma(k) /2\pi = \sigma(k)/2$.} which is yet another way to see that the relation between initial positions of particles with momentum $k$ and  momentum distribution, Eq.~(22) in the main text.

Practically, to plot the trajectories, it is more convenient to obtain from (\ref{xtauGprime})
\begin{align}
    \tau &= -\ii \frac{G'-\bar{G}'}{\varepsilon+\bar{\varepsilon}'}\,,
 \label{tauparam}\\
    x &= \frac{G'\bar\varepsilon'+\bar G'\varepsilon'}{\varepsilon'+\bar\varepsilon'}
 \label{xparam}
\end{align}
and then
\begin{align}
    n = \frac{1}{\pi}\Re \chi = \frac{1}{\pi}\Re (k x-G+i\varepsilon\tau) = \frac{1}{\pi}\Re \chi = \frac{1}{\pi}\Re (k \frac{G'\bar\varepsilon'+\bar G'\varepsilon'}{\varepsilon'+\bar\varepsilon'}-G+\varepsilon \frac{G'-\bar{G}'}{\varepsilon+\bar{\varepsilon}'})\,.
\end{align}
The latter can be rewritten as
\begin{align}
    n = \frac{1}{\pi}\Re \left[kG'-G - (k\varepsilon'-\varepsilon)\frac{G'-\bar G'}{\varepsilon'+\bar\varepsilon'}\right]\,.
 \label{nparam}
\end{align}
The equations (\ref{tauparam},\ref{xparam},\ref{nparam}) give the function $n(x,\tau)$ in a parametric form with two real parameters $\pi \nu$ and $\upsilon$ so that $k,\bar{k}=\pi \nu\pm\ii \upsilon $.
The configurations of the polymers 
are obtained 
as the lines given by $n=const$ with 
\begin{align}
    G(k) &= \frac{1}{2}\int^k \dd k'\,\sqrt{\frac{\sin^2 \frac{k'}{2}-\sin^2\frac{k_0}{2}}{\sin^2 \frac{k'}{2}-\sin^2\frac{k_F}{2}}} \notag\\
    &= \frac{2(1-a^2)}{\sqrt{b^2-a^2}}\left(\Pi\left(\left.\frac{b^2-1}{b^2-a^2}\right\vert\frac{a^2(b^2-1)}{b^2-a^2}\right) -\Pi\left(\left.\frac{b^2-1}{b^2-a^2};\phi\right\vert\frac{a^2(b^2-1)}{b^2-a^2}\right)\right)
\end{align}
given in terms of the  incomplete and complete elliptic functions $\Pi(n;\phi|m)$, $\Pi(n|m)$ with $a=\sin\frac{k_0}{2}$, $b=\sin\frac{k_F}{2}$ and 
\begin{align}
    \phi= \arcsin\frac{\sqrt{1-u^2}}{\sqrt{u^2-a^2}}\frac{\sqrt{b^2-a^2}}{\sqrt{1-b^2}}\, ,\qquad u=\sin\frac{k}{2}\, .
\end{align}

\section*{References}


\bibliographystyle{apsrev4-1}

\bibliography{FCS_draft/FCS}